\documentclass[11pt,a4paper]{article} 
\usepackage{jcappub} 
\usepackage{natbib}

\newcommand{\Rs}{R_{\rm sub}}
\newcommand{\Rt}{R_{\rm tot}}

\newcommand{\vsh}{V_{\rm sh}}
\newcommand{\rsh}{R_{\rm sh}}
\newcommand{\ud}{{\rm d}}

\newcommand{\Mej}{M_{\rm ej}}
\newcommand{\Fesc}{\mathcal{F}_{\rm esc}}
\newcommand{\rxj}{RX J1713.7-3946}
\newcommand{\gr}{$\gamma$-ray}
\newcommand{\degK}{~{\rm K}}
\newcommand{\p}{$\pm$}
\newcommand{\geff}{\gamma_{\rm eff}}
\newcommand{\Ms}{M_{\odot}}
\newcommand{\cmt}[1]{}

\title{Understanding hadronic gamma-ray emission from supernova remnants} 

\author{Caprioli Damiano} 

\affiliation{INAF - Osservatorio Astrofisico di Arcetri, Largo E. Fermi, 5, Firenze, Italy} 

\emailAdd{caprioli@arcetri.astro.it} 

\abstract{
We aim to test the plausibility of a theoretical framework in which the gamma-ray emission detected from supernova remnants may be of hadronic origin, i.e., due to the decay of neutral pions produced in nuclear collisions involving relativistic nuclei. 
In particular, we investigate the effects induced by magnetic field amplification on the expected particle spectra, outlining a phenomenological scenario consistent with both the underlying Physics and the larger and larger amount of observational data provided by the present generation of gamma experiments, which seem to indicate rather steep spectra for the accelerated particles.
In addition, in order to study to study how pre-supernova winds might affect the expected emission in this class of sources, the time-dependent gamma-ray luminosity of a remnant with a massive progenitor is worked out. Solid points and limitations of the proposed scenario are finally discussed in a critical way.
}
 
\begin{document} 
\maketitle 

\section{Supernova remnants, cosmic rays and gamma-ray emission}\label{sec:intro}
The present generation of \gr~telescopes, both in the GeV (Fermi, AGILE) and in the TeV band (HESS, VERITAS, MAGIC, CANGAROO, MILAGRO,...) is providing us with an unprecedented wealth of detections and observations of \gr-bright supernova remnants (SNRs).
These objects have been considered for more than 70 years the main sources of Galactic cosmic rays (CRs) \cite{baade-zwicky34}, but scientists are still looking for a clear-cut evidence of hadron acceleration in these environments. 
Such a \emph{smoking gun} was predicted to be the emission of \gr s due to the decay of neutral pions produced in nuclear collisions between accelerated particles and the background gas \cite{dav94}. 
The most promising place to look at has been considered for some time \rxj~\cite{enomoto+02,Fermi1713}, the prototype of a \gr-bright SNR.

Its TeV emission (between $\sim$100 GeV and $\sim$10 TeV) has been extensively studied by several groups \cite{bv06,gio+09,za10,ellison+10,fang+09,casanova+10,plaga08,yuan+10,Fang+11}, who have investigated two different scenarios: besides the \emph{hadronic} scenario outlined above, also the inverse Compton scattering of the relativistic electrons responsible for the X-ray synchrotron emission may produce a similar signature in the TeV band (\emph{leptonic scenario}).
Each scenario shows strong points and apparent shortcomings, with theorists waiting for Fermi's advent in order to discriminate between them \cite{bv06,gio+09,za10,ellison+10,fang+09,casanova+10,plaga08,yuan+10,Fang+11}. 
The \gr~ spectrum between $\sim$100 MeV and $\sim$10 GeV (hereafter the GeV band) is in fact expected to be significantly flatter in the leptonic scenario with respect to the hadronic one. 
Very recently Fermi-LAT eventually showed with unprecedented resolution the morphology of the \gr~emission in the region around \rxj, favouring a leptonic rather than a hadronic emission from the SNR shell and eventually undermining physicists' hope to have found the smoking-gun for hadron acceleration in SNRs \cite{Fermi1713}.

We start our analysis by building up the list of Galactic SNRs detected in \gr s, followed by the list of other possible candidates (table \ref{tab:SNR}).
This latter class of objects is made of sources whose association with a SNR is plausible but not fully confirmed because of the tangled morphology of the emitting region and/or because of the ``pollution'' due to other potential \gr~sources, like pulsars and pulsar wind nebulae. 
Generally speaking, a clear detection of a SNR \emph{shell} is an exceptional fact (probably the best example is Vela Jr, with SN1006 also showing a sharp bilateral emission consistent with its non-thermal X-ray morphology).
In addition, it is necessary to remember that in many cases the angular resolution of the telescopes does not allow for a detailed morphological study of extended sources. 
The association of the \gr~emission with molecular clouds (MCs) is also very common: in table 2 of ref.~\cite{jiang+10} the reader can find a thorough collection of Galactic SNRs known to be in physical contact with MCs, i.e.\ showing evidences of interaction between the forward shock and the MC. 
Very generally, also the \gr~emission detected from sources not listed in \cite{jiang+10} may be (and has often been interpreted to be) due to MCs surrounding the SNR rather than to the shell itself. 

In the present work we consider only particles accelerated at SNR forward shocks, neglecting both the difficulties related to particle acceleration in dense, partially neutral, environments and to the possibility for the emission to come from MCs illuminated by CRs escaping the SNR. We will comment more on these points below, in section \ref{sec:cav}.
This approach is motivated by the fact that basically \emph{all} of the \gr-bright SNRs show quite similar properties, and in particular energy spectra steeper than $E^{-2}$, independently from their association with MCs. 
Within a hadronic scenario for the \gr~emission, the photon spectrum above $\sim$100 MeV has to be parallel to the proton one, therefore the observational evidences reported in table~\ref{tab:SNR} can put a strict constraint on the physical mechanisms responsible for hadron acceleration in this class of sources.

The success of the so-called \emph{SNR paradigm} for the origin of Galactic CRs, besides the energetic argument already put forward by Baade and Zwicky \cite{baade-zwicky34}, mostly relies on the generality of the spectrum of accelerated particles produced via Fermi's first order mechanism at strong shocks. 
In fact, during the late '70s several authors independently realized that diffusive shock acceleration (DSA) leads to power-law spectra
\begin{equation}\label{slope}
N(E)\propto E^{-\alpha}\,;\qquad \alpha=\frac{r+2}{r-1}\,,
\end{equation}
whose slope does not depend on the microphysical details of the scattering process but only on the shock compression ratio $r$, i.e.\ the ratio between the density of the hotter, shocked plasma and of the colder, unperturbed one \cite{axford+77,krymskii77,ALS78, blandford-ostriker78, bell78a, bell78b}.
In particular, for strong shocks (namely with sonic Mach number much larger than 1) we have
\begin{equation}
r=\frac{\gamma+1}{\gamma-1}\,,
\end{equation}
and hence, for standard monoatomic gas with adiabatic index $\gamma=5/3$, we get $N(E)\propto E^{-2}$.
Such a power-law spectrum is particularly appealing since it is in decent agreement both with the multi-wavelentgh observations of SNRs and, once corrected for propagation in the Milky Way, with the diffuse spectrum of Galactic CRs measured at Earth as well.   

Nevertheless, as soon as the first quantitative calculations about DSA efficiency were carried out, people realized that such a process may channel a large fraction (even more than 90\%) of the fluid ram pressure into CRs. 
In such a scenario, accelerated particles can no longer be viewed as \emph{test-particles} and the CR population has to be treated as an additional component entering the hydrodynamical equations for conservation of mass, momentum and energy \cite{drury-volk81a,drury-volk81b}. 
This \emph{two-fluids} approach to the non-linear theory of DSA (NLDSA) has soon been followed by \emph{kinetic} approaches to the problem, in which also the information about the momentum distribution of accelerated particles is retained.
In this case the equations of the shock hydrodynamics, modified by the inclusion of CRs, are solved along with a description of the CR transport, typically accounted for via a diffusion-convection equation or a numerical (Monte Carlo) treatment. 
Comprehensive reviews on CR-modified shocks can be found in refs.~\cite{drury83,blandford-eichler87,jones-ellison91,malkov-drury01}, while in ref.~\cite{comparison} different kinetic approaches to NLDSA (semi-analytical, numerical and Monte Carlo) are summarized and compared.

The most important prediction of any NLDSA theory is that the pressure in accelerated particles around the shock leads to the formation of a \emph{precursor} in which the upstream fluid is slowed down and compressed.
As a consequence, the proper shock becomes weaker (it is in fact referred to as \emph{subshock}) and provides a reduced heating of the downstream plasma. 
Accelerated particles diffusing in such a modified fluid profile feel different (averaged) compression ratios and, since high-energy particles have diffusion lengths larger than low-energy ones, the overall spectrum can no longer be a simple power-law.
High-energy particles probe in fact the whole precursor, and hence feel the total compression ratio $\Rt>4$, while low-energy particles are confined close the subshock, whose compression ratio is instead $\Rs<4$. 
As a consequence of the intrinsic property of Fermi's first order mechanism --- see eq.~\eqref{slope} --- the resulting spectrum is predicted to be concave, i.e.\ steeper (flatter) than $E^{-2}$ at low (high) energies.

This effect for high-energy particles (above some GeV, as a rule of thumb) can be also accounted for as a pure hydrodynamical effect: the contribution to pressure and energy by relativistic particles (whose adiabatic index is $\gamma=4/3$) actually makes the total (gas + CRs) fluid more compressible.
The global effective adiabatic index can be written (see e.g.\ \cite{chevalier83,blondin-ellison01}) as:
\begin{equation}\label{geff}
\geff=\frac{1}{3}\frac{5+3\xi_{cr}}{1+\xi_{cr}}\,,
\end{equation}
where $\xi_{cr}$ is the fraction of the bulk pressure converted in accelerated particles at the shock.
Since $4/3\le \gamma_{\rm eff}\le 5/3$, we have $4\le r_{\rm eff}\le 7$ and finally $1.5\le\alpha\le 2$: the more efficient the acceleration, the flatter the spectrum of the accelerated particles.
Such a simple trend is recovered in all of the NLDSA models and becomes more and more marked if also particle escape is accounted for: in this case, in fact, the shock behaves as partially radiative and the compression ratio might become much larger than 7, eventually leading to spectra as flat as $\sim E^{-1}$ at the highest energies (see e.g.\ \cite{malkov-drury01,ab06}).

Since the spectrum of hadronic \gr s maps the spectrum of accelerated particles  \emph{unequivocally}, concave photon spectra would naturally represent the smoking gun for very efficient hadron acceleration in SNRs. 
In addition, NLDSA naturally predicts pretty hard spectra at large energies, so that hadronic \gr~emission in the TeV range should typically show a spectrum harder than $E^{-2}$.
Neither the former, stronger evidence, nor the latter, more general, one finds any support in the observations summarized in table~\ref{tab:SNR}: does this fact imply that the \gr~emission can not be of hadronic origin? Or that SNRs are not efficient CR accelerators? Or, even, both of them?

In the following sections we show how including an additional piece of information, namely the amplification of magnetic field due to plasma instabilities excited by the streaming of relativistic particles, it is possible to outline a physically consistent scenario in which SNRs are efficient factories of Galactic CRs and the observed \gr~emission may nevertheless be explained within a hadronic scenario. 
 
In section \ref{sec:CR} we introduce a phenomenological model accounting for particle acceleration and magnetic field amplification at shocks. 
In section \ref{sec:hydro} we outline a quite general treatment of the SNR hydrodynamical evolution, also allowing for the presence of pre-SN winds contributing to create a non-trivial circumstellar environment the shock propagates into. 
We then consider in section \ref{sec:comparison} the case of remnant with a massive progenitor, in order to illustrate the effects induced by the winds launched during pre-SN stages on the time evolution of the \gr~emission, to some extent generalizing the pioneering work in ref.~\cite{dav94}.
In particular, such a choice allows us non only to assess the fundamental role of magnetic field amplification in producing rather steep spectra of accelerated particles, but also to argue that the age distribution of \gr-bright SNRs correlates better with core-collapse SNRs rather than with type Ia SNRs expanding into homogeneous environments (section \ref{sec:evo}).
Our findings are finally commented in section \ref{sec:cav}, where we critically discuss the modeling of magnetic field amplification and the possible role of MCs often surrounding SNRs.
We conclude in section \ref{sec:conclusion}. 

\begin{table}\label{tab:SNR}
\centering
  \begin{tabular}{@{}c@{}c@{}c@{}cc@{ }c@{}c@{}}
    %\hline
    \hline
    \multicolumn{6}{c}{\gr-bright Supernova Remnants} \\
    \hline
  %  \hline
    Coordinates & SNR         & Age(kyr) & $\alpha_{GeV}$ & $\alpha_{TeV}$ & Refs. \\
    \hline
    6.4$-$0.1   & W28 North   & 35--45   & 2.09\p0.08     & 2.66\p0.27    & \cite{W28Fermi,W28AGILE,W28HESS,cg08,jiang+10}\\ 
	            & W28 A,B,C   & 35--45   & 2.19\p0.14     & 2.50\p0.20    & \cite{W28Fermi,W28AGILE,jiang+10}\\  
    8.7$-$0.1   & W30	      & 10--50    & 2.4\p0.07      & 2.72\p0.06    & \cite{cs10,aha+06,W30CANGAROO,0FGL,jiang+10}\\
    31.9$+$0.0  & 3C 391      & 4        & 2.33\p0.11     & ¥             & \cite{cs10}  \\ 
    34.7$-$0.4  & W44         & 15       & 2.06\p0.03     &               & \cite{W44Fermi,jiang+10}  \\ 
    43.3$-$0.2  & W49B        & 1--4     & 2.18\p0.04     & 3.1\p0.3      & \cite{W49BFermi} \\
	49.2$-$0.7  & W51C        & 10--20   & 1.7\p0.3       & 		          & \cite{W51CFermi,W51CHESS,MILAGRO09,jiang+10}\ \\ 
    106.3$+$2.7 &             & 10       &                & 2.29\p0.33    & \cite{G106VERITAS,cg08,jiang+10} \\
	119.5$-$2.1 & Cas A       & 0.330    & 2.01\p0.1      & 2.3\p0.2      & \cite{CasAFermi,CasAVERITAS,CasAMAGIC,jiang+10} \\
	120.1$+$1.4 & Tycho       & 0.438    & 2.3\p0.1	& 1.95\p0.50  & \cite{TychoFermi,TychoVER,jiang+10}\\
	189.1$+$3.0 & IC443       & 30       & 1.93\p0.03     & 3.05\p0.40    & \cite{IC443Fermi,IC443AGILE,IC443MAGIC,IC443VERITAS,MILAGRO09,cg08,jiang+10,wang-scoville92} \\
	205.5$+$0.5 & Monoceros   & 30--150  &                & 2.53\p0.26    & \cite{MonocerosHESS,cg08,jiang+10} \\
	266.2$-$1.2 & Vela Jr.    & 0.6--4   &                & 2.24\p0.04    & \cite{VelaCANGAROO,VelaHESS}  \\
	315.4$+$2.3 & RCW 86      & 1.825    &                & 2.41\p0.16    & \cite{RCW86HESS} \\
	327.6$+$14.6& SN1006 NE   & 1.004    &                & 2.54\p0.15    & \cite{SN1006HESS} \\
	            & SN1006 SW   & 1.004    &                & 2.34\p0.22    & \cite{SN1006HESS} \\
	347.3$-$0.5 & \rxj        & 1.6      & 1.5\p0.1       & 2.04\p0.04    & \cite{RXJFermi,Fermi1713,RXJ1713HESS,uchiyama+07,RXJCANGAROO} \\
	348.5$+$0.1 & CTB 37A     & 1.617    & 2.19\p0.07     & 2.30\p0.13    & \cite{cs10,CTB37HESS,jiang+10} \\
	348.7$+$0.3 & CTB 37B     & 2.7--4.9 &                & 2.65\p0.19    & \cite{CTB37HESS} \\
	349.7$+$0.2 &             & 2.8      & 2.10\p0.11     &               & \cite{cs10} \\
	353.6$-$0.7 & HESS J1731-347 & 27    &                & 2.26\p0.10    & \cite{Tian+10,HESS1731} \\
	\hline
%	\hline
	\multicolumn{6}{c}{Other possible candidates} \\
	\hline
	%\hline
    %Coordinates & SNR         & Age(kyr) & $\alpha_{GeV}$ & $\alpha_{TeV}$ & Refs. \\
	%\hline
	0.0$+$0.0   & SGR A East    & 8      & $\sim$2.2      & ¥             & \cite{aha+06,fm05,jiang+10}\\ 
	12.8$-$0.0  & HESS J1813-178& 0.3--25&                & 2.09\p0.08    & \cite{aha+06,0FGL} \\
	21.5$-$0.9  & HESS J1833-105& 0.8--1 &                & 2.08\p0.22    & \cite{dja+08} \\
	23.3$-$0.3  & W41           & 60--100&                & 2.45\p0.16    & \cite{W41MAGIC,aha+06,0FGL,jiang+10}  \\ 
	27.8$+$0.6  &               & 35--55 &                &               & \cite{cg08} \\
	28.8$+$1.5  &               &  32    &                &               & \cite{cg08} \\
	29.7$-$0.3  & Kes 75        &0.7--0.8&                & 2.26\p0.15    & \cite{dja+08} \\
	35.6$-$0.4  & HESS J1858+020& 30     &                & 2.2\p0.1      & \cite{Paron+11,HESS1731} \\
	40.5$-$0.5  & HESS J1908+063& 20--40 &                & 2.08\p0.10    & \cite{dja+08,jiang+10,G405} \\
	54.1$+$0.3  &               &	2.9  &                & 2.3\p0.3      & \cite{wak+10} \\
	65.1$+$0.6  & 0FGL J1954.4+2838 &4--14 &              &               & \cite{0FGL,tl06,MILAGRO09}\\
	78.2$+$2.1  & $\gamma$ Cygni&    5   &                &               & \cite{cg08,weinstein09,jiang+10}\\
	119.5$+$10.2& CTA 1         & 13--17 &                &               & \cite{cg08} \\
	132.7$+$1.3 & HB3           & 30     &                &               & \cite{cg08,jiang+10}\\
	296.4$-$9.45?& HESS J1507-662 & 1?   &		      & 2.24\p0.16    & \cite{HESSJ1507,tibolla09}\\
	338.3$+$0.0 & HESS J1640+465&  20--40&                & 2.42\p0.14    & \cite{aha+06} \\
	343.0$-$0.6 & RCW 114       &  20    &                &               & \cite{cg08} \\
	359.1$-$0.5 & HESS J1745-303& 20--50 & $\sim$2.17     & 2.71\p0.11    & \cite{HESS1745,jiang+10,tibolla09} \\
    \hline
    \multicolumn{6}{l}{\small $^{a}$ W28N, IC443, W51C, W44 and W49B show evidence of a cut-off around 1--20 GeV.} \\
    \multicolumn{6}{l}{\small $^{b}$ Ref.~\cite{jiang+10} points out SNRs in physical contact with MCs, except Tycho according to ref.~\cite{Tian11}.} \\
	\hline
  \end{tabular}
  \label{tab:label}
  \caption{
 Photon spectral index $\alpha$ inferred in \gr-bright SNRs, both in the GeV and in the TeV bands. 
 Associated systematic errors are typically in the range $\pm$(0.1-0.2).
 Information about SNR nomenclature, ages, distances and associations can be found in the Green's Catalogue (\href{http://www.mrao.cam.ac.uk/surveys/snrs/snrs.data.html}{http://www.mrao.cam.ac.uk/surveys/snrs/snrs.data.html}) and in refs.~\cite{SNRdata1,SNRdata2,SNRdata3}.
  See also \href{http://tevcat.uchicago.edu}{http://tevcat.uchicago.edu} and \href{http://www.mppmu.mpg.de/~rwagner/sources/index.html}{http://www.mppmu.mpg.de/$\sim$rwagner/sources/index.html} for catalogues of known TeV sources.}
\end{table}

\section{Modeling the spectrum of cosmic rays}\label{sec:CR}
During the last decade, X-ray observations of young SNRs have shown evidences of bright narrow rims of non-thermal origin, which have been interpreted as due to synchrotron emission from ultra-relativistic electrons. 
Moreover, from the thickness of these rims it has been possible to put a lower limit for the magnetic field immediately behind the shock, finding evidences of magnetic fields as high as a few hundred $\mu$G, almost two order of magnitudes larger than the standard interstellar field (see e.g. refs.~\cite{Bamba+05,V+05,P+06,uchiyama+07,eriksen+11}).

These amplified magnetic fields are very likely produced by the super-Alfv\'enic streaming of particles accelerated at the shock via plasma instabilities \cite{skilling75a,  skilling75b, skilling75c,bell78a,bell04,bell05,ab08,zpv08,reville+07,ohira+09,mario-anatoly}.
In such a scenario, CRs are able to generate by themselves the magnetic turbulence responsible for their own diffusion in a non-linear interplay which eventually allows accelerated particles to reach energies consistent with the steepening (the \emph{knee}, around $3\times 10^{6}$GeV) observed in the diffuse spectrum of Galactic CRs and to explain it as due to the maximum energy achievable in SNRs \cite{bac07}.
An evidence that magnetic field amplification occurs \emph{upstream} comes from the narrow extension of the X-ray emitting region ahead of the shock in SN1006, as it has been put forward by exploiting very detailed Chandra maps \cite{long+03,morlino+10}.

Two main categories of \emph{streaming instabilities} are usually accounted for: resonant and non-resonant ones, according to the relation between the Larmor radius of the relativistic particles and the wavelength of the excited modes.
Resonant streaming instability has been known since the '70s \cite{skilling75a, skilling75b, skilling75c, bell78a}, while more recently T. Bell worked out a class of short-wavelength modes whose growth may be even faster \cite{bell04, bell05}. 
A kinetic approach to streaming instability can show how resonant and non-resonant modes are two faces of the same coin and that non-resonant modes grow faster than resonant ones only for large shock velocities \cite{ab08}.
For this reason, and also because the role of non-resonant modes in CR diffusion is still to be addressed, we focus our attention only on resonant modes, checking \emph{a posteriori} the possible role of Bell's modes.
In addition, we bypass the time-dependent details of the magnetic turbulence growth, assuming that field amplification proceeds until saturation is achieved.
This should happen when the normalized pressure in the amplified magnetic field is (see \cite{ab06,jumpkin} and references therein)
\begin{equation}\label{saturation}
\frac{P_{B}}{\rho \vsh^{2}}=\frac{B^{2}}{8\pi\rho \vsh^{2}}\approx\frac{\xi_{cr}}{2M_{A}},
\end{equation}
with $M_{A}=\vsh/v_{A}$ the Alfv\'enic Mach number. 
It is important to stress that the saturation above is calculated as a non-linear extrapolation of the result obtained within a quasi-linear theory for magnetic field amplification, where $M_{A}$ is taken in the \emph{background} field, $B_{0}$.
However, since a configuration with perturbations $\delta B\gg B_{0}$ is rather peculiar and likely unstable (also because $\delta\vec{B}\perp\vec B_{0}$), we could \emph{phenomenologically} assume that the Alfv\'enic Mach number entering eq.~\eqref{saturation} should be taken as something closer to $M_{A}(B)$ rather than to $M_{A}(B_{0})$.
This choice would also imply that the Alfv\'en velocity $v_{A}=B/\sqrt{4\pi\rho}$ associated with the turbulence should be calculated in the amplified magnetic field as well.
Finally, since CRs actually scatter against magnetic irregularities rather than against the fluid itself, when $v_{A}$ is not negligible with respect to fluid velocity $u$, the compression ratio actually felt by accelerated particles has to be accordingly modified, namely:
\begin{equation}\label{ratios}
r=\frac{u_{1}}{u_{2}}\to \tilde{r}=\frac{u_{1}+v_{A,1}}{u_{2}+v_{A,2}},
\end{equation}
where subscripts 1 and 2 refer to pre- and post-shock quantities, respectively.
Since Alfv\'en waves are generated upstream by streaming instability, they must travel against the fluid (in direction opposite to the CR pressure gradient), while it seems reasonable to assume isotropic turbulence downstream, which implies $v_{A,2}=0$, on average.  
Moreover, when CR acceleration is efficient $U_{1}=u_{1}/V_{sh}\simeq 1-\xi_{cr}$ and the equation for streaming instability saturation reads (see eq.~42 of \cite{jumpkin})
\begin{equation}\label{saturationSI}
 \frac{P_{B,1}}{\rho_{0}\vsh^{2}}=\frac{B_{1}^{2}}{8\pi\rho_{0}\vsh^{2}}=U_{1}^{-3/2}\frac{1-U_{1}^{2}}{4M_{A,1}}
\end{equation}
from which
\begin{equation}\label{eq:Ma1}
M_{A,1}=\frac{2}{\xi_{cr}}\frac{(1-\xi_{cr})}{2-\xi_{cr}}^{5/2}.
\end{equation}
Finally, dividing numerator and denominator in eq.~\eqref{ratios} by $u_{1}$ and retaining the general formula for the shock compression ratio we get
\begin{equation}\label{eq:r}
\tilde{r}=r \left(1-\frac{1}{M_{A,1}}\right)=\frac{\geff+1}{\geff-1+2/M_{s}^{2}}\left[1-\frac{\xi_{cr}(2-\xi_{cr})}{2(1-\xi_{cr})^{5/2}}\right],
\end{equation}
where $M_{s}=\vsh/c_{s}$ is the sonic Mach number of the shock and $\geff$ is defined in eq.~\eqref{geff}.

\begin{figure}
\centering
	{\includegraphics[width=0.95\textwidth, height=0.44\textheight]{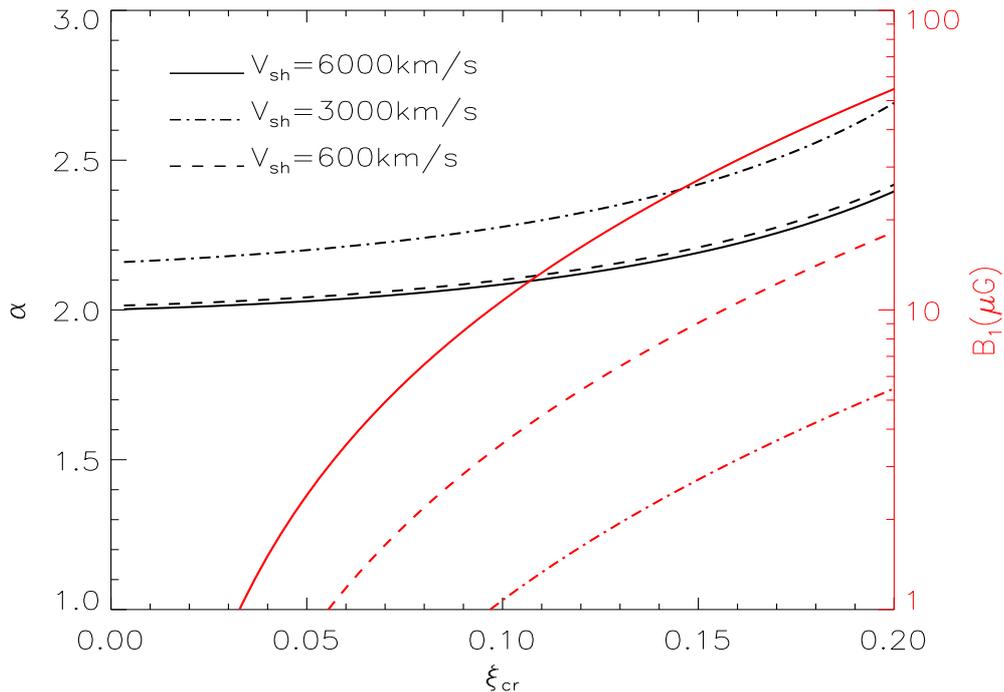}}
	\caption{Slope of the energy spectrum of accelerated particles (dark lines, laft axis) and upstream amplified magnetic field (light lines, right axis) as a function of the CR acceleration efficiency (see eqs.~\ref{saturationSI}--\ref{eq:a}). 
	Temperature and density of the background plasma are fixed in $T_{0}=10^{5}\degK$ and $\rho_{0}=0.1 m_{p}$cm$^{-3}$, while different lines correspond to different shock velocities, as in the legend.}
	\label{eff}
\end{figure}
 
In this work we consider the CR acceleration efficiency $\xi_{cr}$ as given, keeping in mind that, in order to account for the energetics in Galactic CRs, a fraction of about 10$\%$ of the SN kinetic energy has to be channelled into accelerated particles (see e.g. the review in ref.~\cite{hillas05}).
The spectral slope of the accelerated particles is thus simply given, as in eq.~\eqref{slope}, by 
\begin{equation}\label{eq:a}
\alpha=\frac{\tilde{r}+2}{\tilde{r}-1}.
\end{equation}

In figure~\ref{eff} we show the calculated spectral index $\alpha$ (dark lines, left axis) and the related amplified magnetic field upstream of the shock, $B_{1}$ (light lines, right axis) as a function of the CR acceleration efficiency $\xi_{cr}$, taken to vary between zero and 20$\%$.
We fixed temperature and density of the background gas in $T_{0}=10^{5}\degK$ and $\rho_{0}=0.1 m_{p}$cm$^{-3}$, hence the dependence of $\alpha$ on $M_{s}$ is accounted for showing different curves with different shock velocities (as in the legend), corresponding to $M_{s}\simeq 50$,17 and 5, respectively.

From figure~\ref{eff} it is clear that for large Mach numbers a non-negligible CR efficiency is required to produce particle spectra steeper than $E^{-2}$ while, on the other hand, for sufficiently low $M_{s}$ the standard reduction of the compression ratio predicted for non-strong shocks is amplified even further with increasing $\xi_{cr}$. 
Moreover, we notice that the effect of taking the saturation of the instability in the amplified field rather than in the background one completely suppresses the spectral flattening usually predicted by NLDSA theory, so that $r$ never grows beyond 4, neither for large $\xi_{cr}$.
This prediction has however to be taken with a grain of salt, since the simple model outlined here does not account for any concavity in the spectrum of accelerated particles and therefore can be viewed as a heuristic approach which loses its validity whenever a very marked precursors is produced by a very efficient CR acceleration. 
However, since at the moment there are no observational evidences that very large compression ratios are achieved in SNR environments (the most clear example is likely represented by Tycho's SNR, where $4\leq r\leq 7$ is inferred \cite{warren+05}), there are neither consistency issues or observational shortcomings with our working hypotheses.

Finally, we also notice that the amplified magnetic field so produced (right axis in figure~\ref{eff}) are in decent agreement with the ones needed to account for the measured thickness of non-thermal X-ray rims in several young remnants, which require typical downstream magnetic fields $B_{2}\simeq r B_{1}\approx 50-300\mu G$ \cite{V+05,P+06,jumpl}.
Moreover, when instability saturation is given by eq.~\ref{saturationSI}, any dependence on the strength of background magnetic field $B_{0}$ is washed away: the only necessary requirement is the presence of a non-vanshing component of the background field parallel to the CR pressure gradient, i.e. parallel to the shock normal.  

It is worth stressing that the possible role of a finite scattering center velocity has been put forward since when DSA model were first proposed \cite{bell78a}, but only in very recent times, motivated by the discovery of efficient magnetic field amplification in young SNRs, this additional piece of information has been accounted for in order to explain the observed phenomenology (see e.g.~\cite{veb07,zp08b,jumpkin,kang-ryu10}). 
Here we show, for the first time as far as we know, the crucial implication of this phenomenon in the calculation of the hadronic emission from SNRs: the inclusion of such an effect, in fact, at the moment seems to be the only reasonable way to obtain efficient particle acceleration along with steep spectra.  

\section{Modeling the target: non-homogeneous circumstellar environments}\label{sec:hydro} 
Modeling particle acceleration is only half of the problem of understanding the properties of \gr-bright SNRs, the remaining half being the environment they expand into.
This latter ingredient is important because regulates the number density of accelerated particles and provides targets for p--p scattering, as well.
The radiative signature in \gr s due to shock propagation into the dense ($n\simeq 0.1-1{\rm cm}^{-3}$) homogeneous ISM has been already extensively investigated in ref.~\cite{dav94} and in papers based on it. 
Here we want to set up a more general scheme able to account also for complex circumstellar environments, like the ones in which SNRs with massive progenitors expand into.

Core-collapse SNe are the final evolutionary step of very massive stars, which enter the main sequence (Hydrogen burning stage) with more than 5--10 solar masses ($\Ms$) and progressively lose large fractions of their mass in the shape of stellar winds.
In general, the calculation of the exact amount of mass which goes into these winds, and hence of the remainder which goes into the SN ejecta or into the central compact object, depends on many variables, the most important of which are likely the initial mass and the metallicity. 
Let us consider now two major kinds of stellar mass ejection which may occur during pre-SN stages: red-giant and Wolf-Rayet winds.

A single massive star may produce one or even both of them during its life, thus we consider here a quite general case in which a cold, slow and dense red-giant wind is launched, followed by a hot and fast Wolf-Rayet one.
The latter is expected to penetrate through the former, eventually excavating a large, rarefied, cavity around the star, often referred to as a \emph{hot bubble}.
The fine structure of such an interplay is outlined, e.g., in refs.~\cite{chevaliang89,garcia-low1,garcia-low2}, but for our purposes it is worth sketching a rather simpler picture, similar to the one illustrated in figure 3 of ref.~\cite{weaver+77}.

The circumstellar medium is hence taken as spherically symmetrical and as in the following.
The innermost region is occupied by the dense and cold red-giant wind, whose density profile depends on the radial coordinate $r$ as
\begin{equation}
\rho_{w}(r)=\frac{\dot{M}}{4\pi V_{w} r^{2}}=3.5 m_{p}\frac{\dot{M}_{-5,\odot}}{V_{w,6}}\left(\frac{r}{\rm pc}\right)^{-2} {\rm cm}^{-3}\,,
\end{equation}
where $m_{p}$ is the proton mass, $\dot{M}=\dot{M}_{-5,\odot}10^{-5}\Ms$/yr is the mass-loss rate during the wind activity, $V_{w}=V_{w,6}10^{6}$cm/s is the wind velocity; both quantities are normalized to commonly accepted values.
The wind temperature is expected to be $T=10^{4}-10^{5}$\degK~\cite{weaver+77}.
The magnetic field in such a wind is usually thought to be in the shape of a Parker spiral, and therefore exactly perpendicular to the shock normal: this fact may be relevant for our purposes since without a magnetic field component parallel to the CR gradient, the standard streaming instability introduced above should not be very effective. 
On the other hand, it is very likely for the hydro instabilities due to the Wolf-Rayet wind crossing to produce a turbulent (disordered) magnetic field: as a consequence the shock should be predominantly oblique rather than perpendicular and therefore all of the arguments about magnetic field amplification put forward in section \ref{sec:CR} should qualitatively apply to this region as well.  
The radial extension of the wind zone $R_{w}$ is given by the total mass ejected during the red-giant stage, $M_{w}$, i.e.: 
\begin{equation}
M_{w}=\int_{0}^{R_{w}}4\pi\rho_{w}(r){\rm d}r \to R_{w}=1.4 \frac{M_{w,\odot}V_{w,6}}{\dot{M}_{-5,\odot}}{\rm pc}\,.
\end{equation}
Beyond $R_{w}$ we find the hot bubble excavated by the Wolf-Rayet wind, which is usually taken as much more homogeneous and rarefied. 
Fiducial values for temperature and density are $T_{b}\sim10^{6}$\degK ~and $\rho_{b}=0.01 m_{p}{\rm cm}^{-3}$.
Like for the dense wind, the extension of the hot bubble can be determined by the total mass ejected during the Wolf-Rayet stage, namely:
\begin{equation}
M_{b}=\int_{0}^{R_{b}}4\pi\rho_{b}(r){\rm d}r \to R_{b}=10.23 \left(M_{b,\odot}\frac{0.01m_{p}}{\rho_{b}}\right)^{1/3}{\rm pc}\,.
\end{equation}
Phenomena occurring at the transition between the hot bubble and the interstellar medium (ISM) are thoroughly studied in ref.~\cite{weaver+77} but, for our purposes, we can simply assume a sharp discontinuity by posing, for $r>R_{b}$, $T_{0}=10^{4}$\degK, $\rho_{0}/m_{p}=1{\rm cm}^{-3}$.

Massive stars typically lose a large fraction of their initial mass in stellar winds, therefore only a minor fraction goes into the ejecta during the SN explosion. 
In terms of SNR evolution, this fact implies that the remnant of a core-collapse SN enters its Sedov-Taylor stage when its forward shock is still propagating inside the progenitor wind, and precisely at $R_{ST}\simeq 1.4 M_{ej,\odot}$pc for the case above.
For smaller radii (i.e.\ during the ejecta-dominated stage), position and velocity of the forward shock can be calculated with suitable self-similar analytical solutions \cite{TMK99}.
Here we adopt the same recipe as in eq.~15 of ref.~\cite{pz05}, which reads: 
\begin{eqnarray}
 t(\rsh) &\simeq& 99 R_{\rm sh, pc}^{8/7}\left(\frac{\mathcal{E}_{51}^{7/2}V_{w,6}}{\dot{M}_{-5,\odot} M_{\rm ej,\odot}^{5/2}}\right)^{-1/7} {\rm yr} \\
 \vsh(\rsh) &\simeq& 8800 R_{\rm sh, pc}^{-1/7}\left(\frac{\mathcal{E}_{51}^{7/2}V_{w,6}}{\dot{M}_{-5,\odot} M_{\rm ej,\odot}^{5/2}}\right)^{1/7} {\rm km~  s^{-1}}.
\end{eqnarray}

For $r\ge R_{ST}$ the forward shock propagation into the complex circumstellar environment can be, instead, calculated by adopting the so-called \emph{thin-shell approximation}, i.e.\ neglecting the spatial structure inside the shock and deriving position ($\rsh$) and velocity ($\vsh$) of the forward shock by assuming the total mass (ejecta + swept-up) to be concentrated in a thin shell around $\rsh$.
The shock evolution is hence obtained by solving the time-dependent equations for continuity of mass, momentum and energy and reads (see appendix of ref.~\cite{pz05} and ref.~\cite{omy10}):
\begin{eqnarray}
 M(\rsh) &=& \Mej+4\pi\int_{0}^{\rsh}{\rm d}rr^{2} \rho(r); \\
 \mathcal{E}(\rsh) &=& \mathcal{E}_{SN}-4\pi\int_{0}^{\rsh} \ud r r^{2} \Fesc (r); \\
 t(\rsh) &=& \int_{0}^{\rsh}\frac{\ud r}{\vsh(r)};\qquad  \lambda = 6\frac{\geff-1}{\geff+1}\\
 \vsh(\rsh) &=&\frac{\geff+1}{2}\left[\frac{2\lambda }{M^{2}(\rsh)\rsh^{\lambda}}\int_{0}^{\rsh}\ud r r^{\lambda-1}E(r) M(r)\right]^{1/2},
\end{eqnarray}
where $\Mej$ and $\mathcal{E}_{\rm SN}=\mathcal{E}_{51}10^{51}$erg are respectively the mass and the kinetic energy of the SN ejecta.
$\Fesc$ is the flux of energy carried away from the system by escaping CRs, assumed to be negligible during the ejecta-dominated stage (see ref.~\cite{escape} for details). 
This non-adiabatic term adds up to the standard approach of including the CR contribution by adopting a suitable adiabatic index for the fluid, as in eq.~\eqref{geff}, and allows us to test an additional effect of an efficient CR production on the shock dynamics (also see ref.~\cite{EDB04}).
The evolution of the remnant is followed until the end of the Sedov-Taylor stage. 
A generalization of this method able to account for the transition to the radiative stage could be performed following the analytic approach put forward in ref.~\cite{bandiera-petruk04}, but it is beyond the main goal of this paper.

\section{What observations say}\label{sec:comparison}
The mathematical apparatus put forward in the previous sections may be applied to basically any given SNR since it can easily account either for red-giant or Wolf-Rayet wind only, or even describe the evolution of type Ia SNRs expanding in the dense and cold homogeneous ISM.
The CR acceleration and the argument about the role of the amplified magnetic field hold basically independently of the circumstellar environment.

In this section we choose to focus our attention on remnants produced by core-collapse SNe for several reasons: 
1) core-collapse (type Ib/c and type II) SNe are more common than type Ia ones, accounting for about 70--80\% of total Galactic SNe \cite{hl05}; they are therefore more representative for our statistical analysis; 
2) in young SNRs (with strong shocks) the spectral slope depends only on the CR acceleration efficiency, as showed in figure \ref{eff}, therefore the key point concerning the need to explain the steep spectra observed is almost independent of the details of the circumstellar medium;
3) it is indeed interesting to investigate the time evolution of the \gr~ luminosity when the shock propagates into a circumstellar medium significantly modified by the stellar winds launched during pre-SN stages.
This latter aspect is complementary to the analysis already led for a homogeneous circumstellar medium (\cite{dav94} and papers based on it) and may give an important insight into the nature of the progenitors of \gr-bright SNRs. 
In particular, in section \ref{sec:evo} we point out how the observed age distribution of the sources in table \ref{tab:SNR} is more consistent with the one expected from SNRs whose progenitor blew a rarefied hot bubble rather than from SNRs expanding in the unperturbed ISM.

Let us consider a massive star whose total mass in ejecta, red-giant wind and Wolf-Rayet wind (i.e.\ the hot bubble mass) are respectively $\Mej=4\Ms$, $M_{w}=12\Ms$ and $M_{b}=16\Ms$. 
Let us also take standard values $\mathcal{E}_{51}=V_{w,6}=\dot{M}_{-5,\odot}=1$.
In addition, CR acceleration efficiency is fixed in $\xi_{cr}=0.1$ and the instantaneous flux of energy carried away by escaping CRs $\Fesc$ is self-consistently calculated according to eq.\ 24 in ref.~\cite{escape}.

\begin{figure}
		\includegraphics[width=\textwidth]{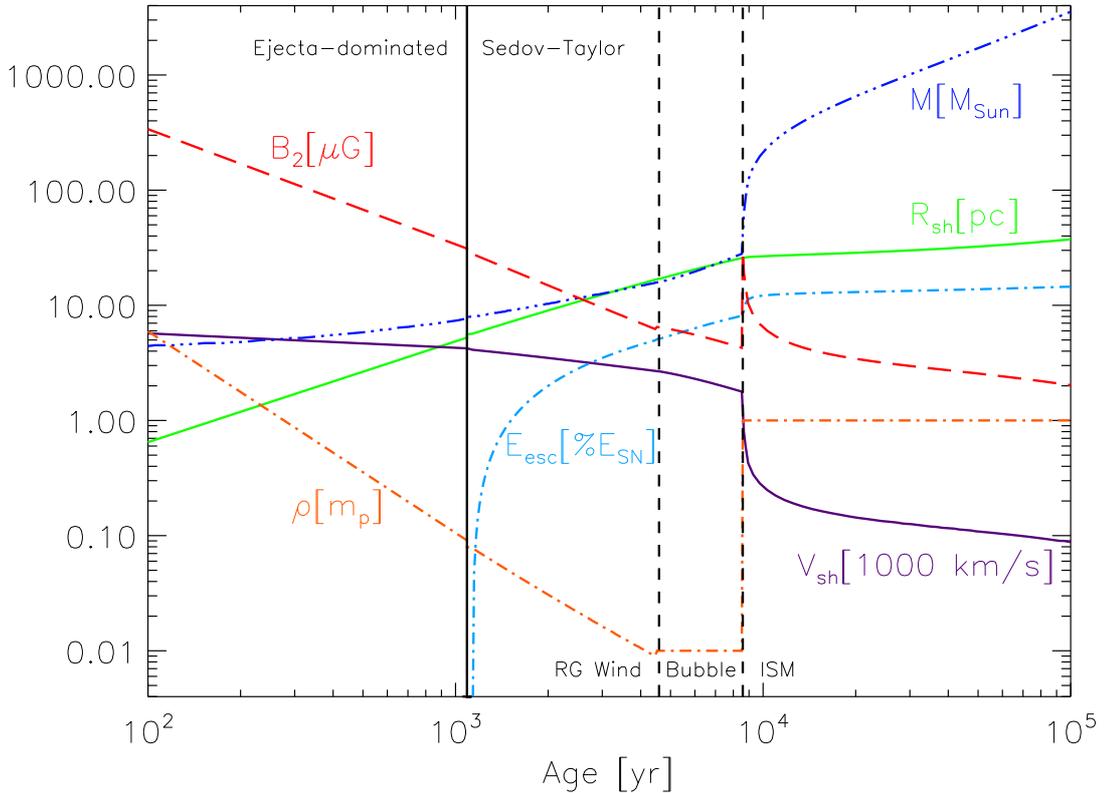}
	\caption{Time evolution of relevant physical quantities for a fixed CR acceleration efficiency $\xi_{cr}=0.1$. 
	The vertical solid line indicates the transition between ejecta-dominated and Sedov-Taylor stages, while vertical dashed lines, from left to right, mark the boundaries of wind zone, hot bubble and ISM, as in the labels (see also the description in section \ref{sec:hydro}).}
	\label{Hydro}
\end{figure}

The evolution of relevant physical quantities is depicted in figure \ref{Hydro}.
The vertical solid line marks the boundary between the ejecta-dominated and the Sedov-Taylor stages, while vertical dashed lines indicate the SNR evolutionary stages in terms of the medium the forward shock propagate into: from inside to outside we have the red-giant wind, the hot bubble and finally the ordinary ISM, as described above.

\begin{figure}
\centering
	{\includegraphics[width=0.95\textwidth, height=0.44\textheight]{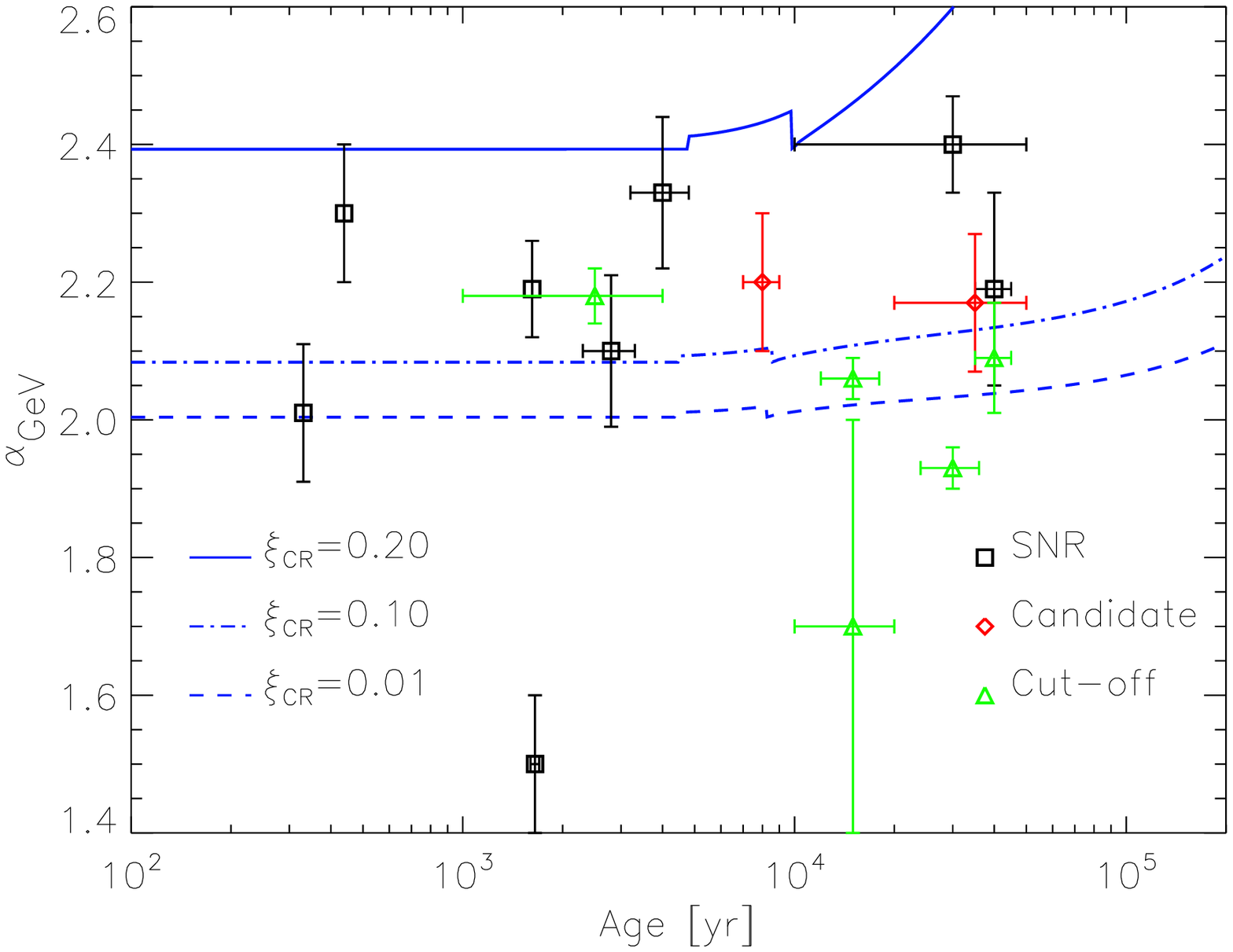}}
	{\includegraphics[width=0.95\textwidth, height=0.44\textheight]{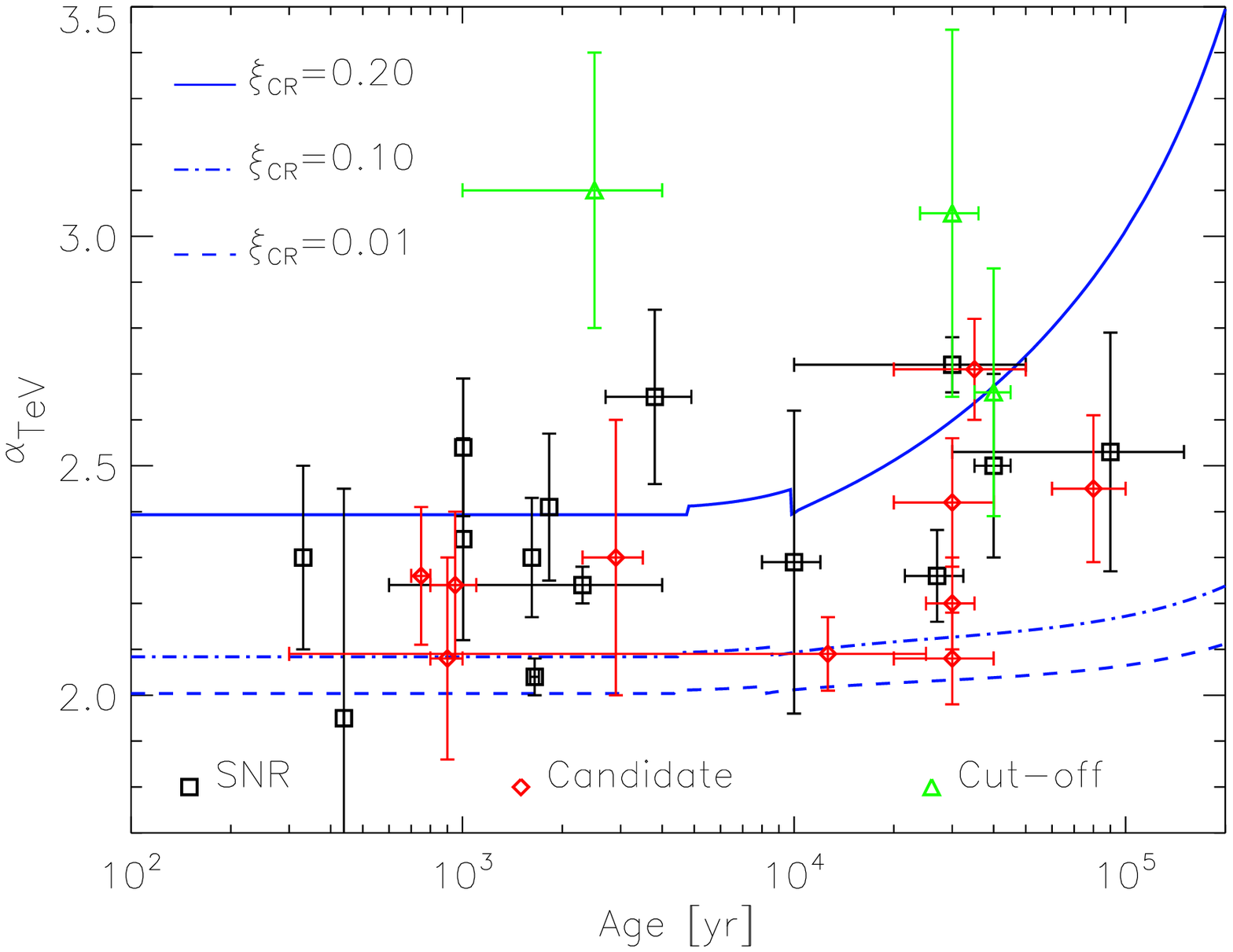}}
	\caption{\gr-bright SNRs detected in the GeV (top panel) and in the TeV (bottom panel) band as in table \ref{tab:SNR}. 
	Different lines correspond to the CR spectral slope as a function of time, for different acceleration efficiencies as in the legend.}
	\label{gevtev}
\end{figure}

It is interesting to notice that the standard adiabatic solution for the shock position and velocity is recovered until the amount of energy carried away by escaping particles (around 10\% of $\mathcal{E}_{SN}$) begins to play a non-neglibile role in the shock dynamics, as showed by the steepening of the velocity curve after $\sim 3000$ yr. 
The SNR radius eventually almost stands when the forward shock encounters the dense and cold ISM, slowing down quite abruptly.
Another interesting evolution worth noticing is the one of the downstream magnetic field ($B_{2}$), which is efficiently amplified by CR streaming instability during early stages and then drops to its typical interstellar value (a few $\mu$G) as the shock velocity decreases. 
Such a trend is in qualitatively good agreement with the observational evidence that only young SNRs are very bright in non-thermal X-rays, and it is also in quantitative agreement with the strength --- a few hundreds $\mu$G --- of the downstream magnetic fields inferred in young shell-type SNRs \cite{Bamba+05,V+05,P+06,uchiyama+07,eriksen+11}.

In figure \ref{gevtev} the predictions for the slope of the CR spectrum, $\alpha$, (and therefore for the photon index above $\sim$100 MeV in a hadronic scenario) are plotted as a function of the SNR age for three different acceleration efficiencies: $\xi_{cr}=0.01, 0.1$ and 0.2, as in the legend.
The upper panel refers to observation by \gr~satellites in the 100 MeV--50 GeV energy range, while the bottom panel illustrates data from Cherenkov telescopes in the region between 50 GeV and 50 TeV.
Source data collected in table \ref{tab:SNR} are divided into three categories: SNRs (black squares), SNR candidates (red diamonds) and finally sources showing evidences for a cut-off around 1--10 GeV (green triangles).

The most striking result of the present calculation is that magnetic field amplification due to CR streaming instability may help to explain the spectral slopes inferred by \gr~observation of SNRs, provided the acceleration efficiency to be as large as 10-20\%.
In fact most of the data points, both in the GeV and in the TeV range, fall above the line corresponding to the inefficient case ($\xi_{cr}=0.01$, dashed line) and lie between the moderately efficient case ($\xi_{cr}=0.1$, dot-dashed line) and the efficient one ($\xi_{cr}=0.2$, solid line). 
More precisely, the dependence of the spectral slope on the CR acceleration efficiency is rather strong above $\xi_{cr}\sim 0.1$, so that efficiencies larger than 20\% are never required.
This fact is particularly appealing also because NLDSA has widely shown how intrinsically difficult it is to produce spectra much steeper than $E^{-2}$ and, at the same time, to achieve large acceleration efficiencies.
It is important to stress that results here have been worked out not within a fully non-linear theory of DSA, but rather within a hybrid approach which accounts in a decent way for the CR feedback on the SNR evolution.
Such a treatment neither retains any kinetic information about the spectral concavity, nor evaluates in a self-consistent way the actual position of the cut-off in the proton spectrum.
In this respect, it is indeed useful to recall that both the preliminary NLDSA calculations in section 5.1 of ref.~\cite{jumpkin} and the more complete ones outlined in ref.~\cite{Texas10} demonstrate that it is possible to get a moderately efficient scenario consistent with CR spectra significantly steeper than the test-particle prediction by accounting for an effective scattering center velocity, as in section \ref{sec:CR}.
As a general consideration, as long as we are dealing with efficiencies of order $\sim 10\%$, non-linear effects like a strong upstream precursor implying a strong spectral concavity can be safely neglected (e.g.\ \cite{kang-ryu10}), in turn justifying \emph{a posteriori} the assumptions made in the present work (also see section \ref{sec:Bmodel} for further comments on this point).

The simple approach proposed here includes many physical ingredients which allow us not only to study in a consistent way the remnant evolution, but also to have an estimate of the properties of the expected \gr-emission. 
A prediction of this model is that, in young SNRs, the CR spectral index may depend only on the CR acceleration efficiency.
In fact, as long as both sonic and Alfv\'enic Mach numbers are large, $\alpha$ depends only on $M_{A,1}$ which is, in turn, a function of $\xi_{cr}$ only (eq.~\ref{eq:Ma1}). 
It is worth stressing that this fact is a direct consequence of having assumed the Alfv\'en velocity relevant for streaming instability saturation to be the one in the amplified magnetic field, and not in the background one. 
 
When $M_{A}$ decreases because of the low density (as in the hot bubble) or when the shock slows down because of the swept-up mass, the spectral slope gradually increases: this is the reason why middle-age and old SNRs are expected to show rather steeper spectra.
In addition, also the energy carried away by escaping CRs may be relevant for the determination of the spectral slope: lines corresponding to different $\xi_{cr}$ in figure \ref{gevtev} are in fact parallel to each other until $\mathcal{E}_{esc}$ becomes a non-negligible fraction of $\mathcal{E}_{SN}$. 
This energy depletion due to CR escape might lead, in principle, to arbitrarily steep spectra for old SNRs, as suggested by solid lines in figure \ref{gevtev}, and/or to an early death of the remnant in terms of its non-thermal activity.

Some TeV data points seem to fall quite outside the region consistent with our predictions, but most of these correspond to sources showing evidence for a cut-off in the \gr-emission around 1--10 GeV (green triangles in figure \ref{gevtev}).
For these sources the physical scenario depicted above does not apply because their TeV spectrum may resemble more an exponential tail rather than an ordinary power-law.
Such a warning also applies, to minor extent, to basically all the TeV data points since, apart from few well-studied SNRs showing clear-cut evidences for a cut-off around some TeV, the superposition of multiple sources and/or the low statistics do not allow us to distinguish between a cut-off spectrum or a steeper power-law.
In this respect, the wealth of new, high-resolution data obtained with present and future \gr~experiments is expected to provide more and more accurate information about the statistics of SNR \gr~spectra, in turn allowing a better and better comprehension of the relationship among remnant age and both spectral slope and instantaneous maximum energy $E_{max}$ of accelerated hadrons.

An interesting information which could be extracted from \gr~ spectra, when interpreted to be of hadronic origin, could in fact be the distribution of $E_{max}$ as a function of the SNR evolutionary stage. 
While there is a wide consensus on the expected increase of the age-dominated $E_{max}$ during the ejecta-dominated stage (see e.g. refs.~\cite{bac07,lagage-cesarsky83b,lagage-cesarsky83a}), the actual behavior of $E_{max}(t)$ during later stages is still an open question. 
The expected decrease of the magnetic turbulence level suggests a space-limited $E_{max}$, determined by a less and less efficient confinement of high-energy CRs.
Unfortunately, quantitative calculations of the actual magnetic field damping in middle-age SNRs depend on a plethora of plasma processes very difficult to keep under control from first principles \cite{pz03,pz05}.

\subsection{Evolution of the gamma-ray emission}\label{sec:evo}

\begin{figure}
\centering
	{\includegraphics[width=0.95\textwidth]{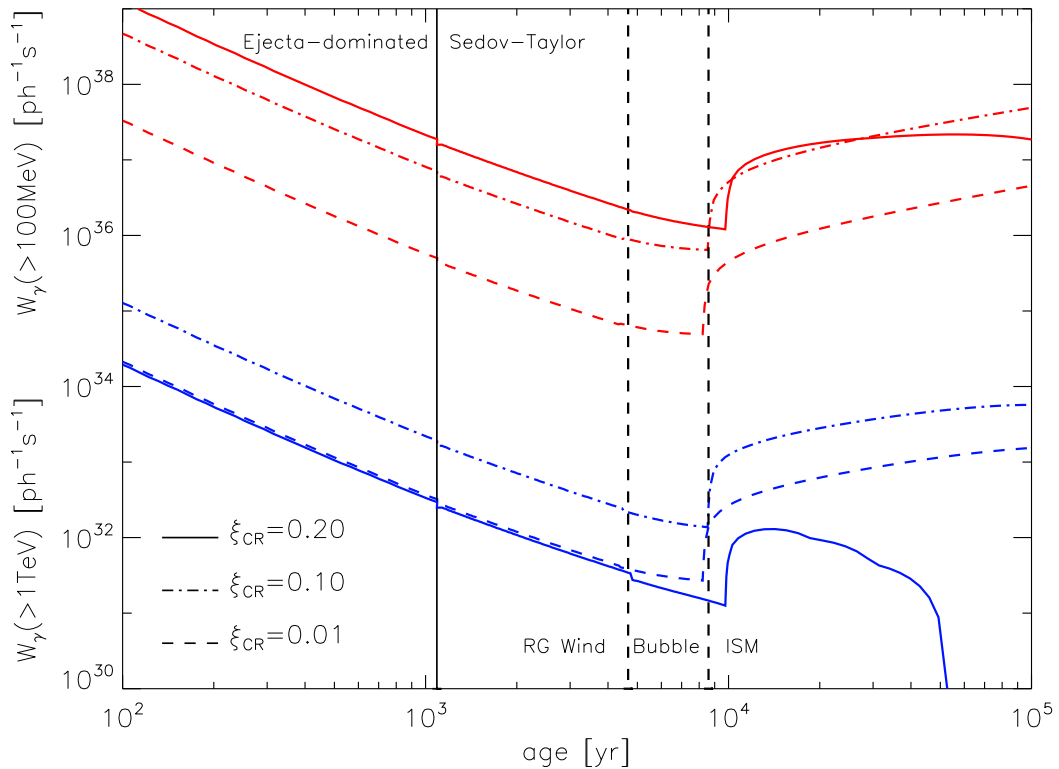}}
	\caption{Time evolution of SNR \gr~luminosity, both in the GeV (upper lines) and in the TeV band (lower lines). 
	Solid, dashed and dot-dashed lines correspond to different CR acceleration efficiency, as in the legend.
	Vertical lines illustrate the evolutionary stages for the case $\xi_{cr}=0.1$, as in figure \ref{Hydro}.}
	\label{power}
\end{figure}

In this section we work out the time evolution of the expected \gr~luminosity by adopting the same techniques of ref.~\cite{dav94}, not in the case of the homogeneous ISM but for the circumstellar profile outlined above, instead.
At any given time, in fact, it is possible to estimate the total number of expected \gr s with energy above $E_{0}$ as
\begin{equation}
W_{\gamma}(>E_{0},t) \simeq q_{\gamma}(E_{0},\alpha) M(t)\frac{\mathcal{E}_{\rm cr}(t)}{\frac{4\pi}{3}\rsh^{3}(t)},
\end{equation}
where $q_{\gamma}(E_{0},\alpha)$ contains the whole information about nuclear interactions and about the shape of the proton spectrum (see table 1 in ref.~\cite{dav94}), $M(t)$ is, as usual, the total swept-up mass and $\mathcal{E}_{\rm cr}(t)$ is the total energy in CRs, i.e.:
\begin{equation}
\mathcal{E}_{\rm cr}(t)=\int_{0}^{t}\ud \tau \frac{1}{2}\xi_{cr}\rho(\tau)\vsh^{3}(\tau)4\pi \rsh^{2}(\tau).
\end{equation}

In figure \ref{power} the total luminosity in photons per second above 100 MeV (upper lines) and above 1 TeV (lower lines) is plotted; different lines correspond to the three CR acceleration efficiencies considered above.
The most interesting feature is that there is a minimum of the emission when the forward shock propagates into the hot bubble, which can easily be interpreted as a consequence of the scarce amount of targets available during this stage.
The total luminosity may be orders of magnitudes larger at earlier and later stages, inside the dense wind and inside the ISM, respectively. 
In passing by, we also notice that in the efficient case ($\xi_{cr}=0.2$, solid line in figure \ref{power}) the impact of the forward shock with the dense ISM occurs at later times with respect to less efficient cases: this fact is a consequence of the 
energy carried away by escaping CRs, which efficiently slows down the shock already during its propagation inside the hot bubble (compare the position of steep rise of the solid curve around $10^{4}$ yr with respect to the vertical dashed line marking the transition between the bubble and the ISM for the case $\xi_{cr}=0.1$, as in figure \ref{Hydro}). 

By looking at the inferred ages of \gr-bright SNRs, it is possible to notice a sort of bimodal distribution: most of the SNRs seem to be either quite young (less than 3000 yr-old) or rather old (more than 10000 yr-old).
The statistics may still be quite low to claim a solid evidence for a dichotomy in the distribution but, on the other hand, we showed that it is quite natural to expect a significant reduction of the hadronic \gr-emission when the shock is propagating inside the hot bubble.
For massive progenitors, this SNR stage should begin around a few thousands yr after the SN explosion and last as much time, depending on the mass-loss rate and total mass ejected during the red-giant and the Wolf-Rayet wind periods.

Since in the Milky Way core-collapse SNe are more frequent by a factor 4-5 \cite{hl05} than type Ia ones, which explode in the homogeneous ISM and are not expected to show a drop in the \gr~luminosity at intermediate stages, it is plausible for the signature of a bimodal distribution to pop up in a representative sample of Galactic \gr-bright SNRs.

On the other hand, SNRs with type Ia progenitors show a peak in the emission around 5000 yr \cite{dav94}, exactly in the less populated region of figure \ref{gevtev}.
Therefore, the age distribution of \gr-bright SNRs listed in table \ref{tab:SNR} is quite at odds with a scenario accounting only for SNRs expanding into the unperturbed ISM and seems to favor the correlation with SNR expanding, for a limited stage of their life, in the rarefied bubble excavated by fast pre-SN winds.
In addition, the presence of dense red-giant winds may help to explain the pretty large number of very young SNRs (say with ages below 3000 yr) detected in \gr s, which would not be accounted for if young SNRs expanded in the hot bubble only.
The amounts of mass going into the different winds in the present example (figure \ref{Hydro}) have hence been chosen in order to match the stage of propagation in the bubble in correspondence with the underpopulated age region in figure \ref{gevtev} (between 3000 and 10000 yr).   

It is however important to stress that core-collapse SNe are typically found in cluster of young stars and are thus often associated with MCs which may significantly enhance the \gr~luminosity by providing extra targets.
If this were the most common case among the objects in table \ref{gevtev}, however, it would not be easy to unequivocally associate the age distribution of \gr-bright SNRs with the nature of their progenitor.

This topic is indeed particularly intriguing since the actual factories of Galactic CRs have been preferentially related either to isolated SNRs exploding in the standard ISM on the basis of the CR chemical composition (see e.g. refs.~\cite{mde97}), or to SN-rich superbubbles \cite{butt09}.
Future \gr~observations might unravel the question, providing an interesting insight into the nature of the medium CR accelerators explode into by identifying possible ``gaps'' in the number of detected sources as a function of their age.

\subsection{GeV/TeV connection}\label{sec:gevtev}
As long as a SNR is observed in a single band, it is very difficult to unambiguously identify the mechanisms responsible for its non-thermal emission and, in fact, every studied SNR shows a \gr~emission that could be (and has been) fitted either within a leptonic or within a hadronic scenario.
The idea behind this point is that a spectrum steeper than $E^{-2}$ can be accounted for either as the result of the decay of neutral pions produced in hadronic interactions or as due to the relativistic bremsstrahlung from accelerated electrons.
In addition, in many cases also the inverse-Compton scattering of relativistic electrons on a suitable photon background can provide a relevant contribution in the same energy region. 
Inverse-Compton photon spectra produced by electrons distributed as a power-law $E^{-\alpha}$ are still power-laws, with spectral index $\beta=(\alpha-1)/2<\alpha$. 
However, when observed in a limited range of energies close to or above the cut-off, also a rather flat inverse-Compton contribution may mimic a steeper spectrum, hence it is very hard to rule out such an emission mechanism even if basically all of the observed spectra are steeper than $E^{-2}$ (see e.g.~\cite{eb07} for a wider discussion). 

One could expect such an ambiguity to be washed away when broadband (i.e.\ from sub-GeV to multi TeV) data were available, because of the information enclosed in the relative normalization of the GeV and TeV spectra.
This makes perfect sense, but at the moment there are only a few SNRs detected in both bands (see table \ref{tab:SNR}): in Cassiopeia A the superposition of relativistic bremsstrahlung and inverse-Compton contributions provides a \gr~spectrum consistent with the data as well as the hadronic scenario \cite{CasAFermi}; for \rxj~ purely leptonic, purely hadronic or also mixed scenarios \cite{za10,yuan+10} have been proposed until only very recently Fermi-LAT observations firmly shifted the paradigm towards the leptonic case.
Finally W51C, W28, W49B and IC443 seem to be representative of a class of middle-age SNRs associated with MCs, whose TeV emission (not detected in W51C, yet) should come from above the cut-off \cite{omy11}.

In this respect, observations in the GeV band of TeV-bright SNRs may be extremely important for assessing the role of SNRs as hadronic accelerators in that, in this range of energies, only relativistic bremsstrahlung and pion decay may produce a photon energy spectrum steeper than $E^{-2}$.
There are in fact a few SNRs, like Tycho and CTB37A, whose GeV-to-TeV \gr~emission shows a uniformly steep spectral index: if relativistic bremsstrahlung were negligible in the GeV band there would be no lepton-induced mechanisms able to account for the broadband emission, therefore strongly supporting a hadronic scenario. 
In order to make such a claim, however, it is necessary to carry out a full non-linear modeling of the multi-wavelength emission, from radio to \gr s, able to self-consistently constrain all the different tiles of the observational mosaic. 
In such an analysis important information may come also by the thermal X-ray emission (both continuous and lines), which can put strong constraints on the density of the circumstellar medium and, in turn, on the level of the hadronic emission (see \cite{ellison+10} for the case of \rxj, where there are no evidence of thermal X-ray emission). 

It is also important to stress that NLDSA calculations may also predict a concave spectrum for the accelerated particles, which has to be typically steeper (harder) at the low (high) energies: such an evidence of concavity would be of great importance in assessing SNRs as very efficient CR accelerators. 
However, since when the Alfv\`en velocity is taken in the amplified magnetic field very large CR efficiencies (larger than $10-20\%$) are neither required nor predicted (see also \cite{Texas10}), it is very likely for the deviation from a straight power-law from GeV to TeV energies to be well inside any measurement error.

In any case, since DSA is charge independent, the electron spectrum has to be parallel to protons' one, therefore the argument about the necessity of some physical mechanism able to account for steep spectra applies also when explaining the \gr~emission as due to relativistic bremsstrahlung, as it has been proposed for Cas A \cite{CasAFermi}.

The case of SNR \rxj~ remains however emblematic in highlighting the difficulties embedded in the study of a given source: assessing the dominance of one scenario over the other required in fact a few years of Fermi-LAT observation.
Nevertheless, it is important to notice how in figure \ref{gevtev} \rxj~lies in a quite peculiar position, showing a GeV spectrum much flatter than the average: it is likely for \rxj~not to be the most representative \gr-bright SNR among the ones listed in table~\ref{tab:SNR}.

It is indeed possible that different emission mechanisms are at work at the same time, but in different spatial regions determined, for instance, by a strongly inhomogeneous circumstellar medium populated with several MCs. Non-spherical models of such complex environments have not been put forward yet, but nevertheless the theoretical framework is already developed enough for testing future, detailed, space-resolved maps of the broadband emission.
In this respect, very interesting information may come in the very next future from systems showing emission both from the proper SNR shell and from close MCs, like for instance in the cases of W28, W51 or \rxj.

\section{\emph{Caveats} and limitations of the present approach}\label{sec:cav}
The simple model put forward in this paper is only meant to provide a first-order quantitative insight into the phenomenology of hadronic \gr-emission from SNRs. 
In this section we discuss the limitations of the present approach and outline some other physical ingredients which may enter a more accurate prediction of \gr~spectra and fluxes expected from SNRs.

\subsection{Magnetic field modeling}\label{sec:Bmodel}
Both the present treatment of magnetic field amplification and the adoption of an effective scattering center velocity are based on a plausible extrapolation in the non-linear regime of the quasi-linear theory, and has therefore to be checked against numerical simulations for assessing the validity of the assumptions made.
We checked \emph{a posteriori} that the inclusion of Bell's non-resonant modes \cite{bell04,bell05}, with their standard saturation $P_{cr}/P_{B}\simeq 2c/\vsh$ \cite{ab08,zpv08,reville+07,ohira+09,mario-anatoly}, does dot change the results above in a sizable way. 
The very reason of this fact is that Bell's mode saturation does not depend on the Alfv\'enic Mach number, i.e.\ magnetic field amplification is not expected to become more and more effective the larger the magnetic field is, while resonant streaming instability   
may saturate to a level which eventually depends only on $\xi_{cr}$ (see eq.~\ref{saturationSI}).
As a consequence, when such an effective saturation is achieved, typical values of $\xi_{cr}$ make excitation of resonant modes so efficient to overwhelm the possible contribution of smaller-wavelength ones.
However, the intrinsic complexity and non-linearity of the interplay between relativistic particles and magnetic fields is such that a deep comprehension of how magnetic field amplification really occurs, and hence of how to model it correctly for observational purposes, might come only from particle-in-cell simulations (see e.g.~\cite{mario-anatoly} and references therein).   

It is fair to stress that, from a quantitative point of view, the recipes put forward in section \ref{sec:CR} have to be regarded as phenomenological tools, in that results might be somewhat different if one assumed different prescriptions either for the wave anisotropy, or for the effective fraction of the amplified magnetic field entering the Alfv\'en velocity \cite{veb06} or even for the wave transport equation, a solution of which is given by \ref{saturationSI}.

In the two-fluid model described in section \ref{sec:CR} we dealt with relatively small CR acceleration efficiencies, hence the resulting shock are only mildly modified.
However, even with almost constant velocity and density in the upstream, particles may in principle feel different compression ratios because of variations in the local Alfv\'en velocity.
A curvature in the spectrum may in turn result from the fact that particles with different momenta feel different averaged Alfv\'en velocities (i.e.\ different magnetic fields) in the precursor. 
Since we assumed the magnetic field to be constant upstream as well, no concavity can be consistently predicted. 
The question may be whether this last assumption is realistic or too much simplistic and therefore whether spectra might be concave. 
Strictly speaking, in fact, magnetic field amplification should be more efficient where the gradients in the CR distribution function are larger, namely in the correspondence of the diffusion length of the particles carrying the most of the energy (around 1 GeV for spectra steeper than $E^{-2}$). 
Particles with larger energies and therefore larger diffusion lengths actually probe regions with lower magnetic field, therefore the spectrum may become flatter and flatter at higher and higher energies.

Some arguments supporting the fact that our two-fluid model can nevertheless be a reasonable approximation of a more complex kinetic model are the following:
1) in the far precursor non-resonant streaming instability excited by escaping CRs may help to provide a substantial magnetic field enhancement, actually reducing the gradient in the magnetic field; 
2) the possible flattening at the highest energies is however controlled by the largest compression factor ($r\approx 4-4.5$) allowed by $\gamma_{eff}$ with $\xi_{cr}\simeq 0.1$, hence the spectral index may vary between 2.2--2.5 and 1.9--2 at most.
Measuring a possible hint of concavity could be very intriguing but present observations, whose systematic + statistical errors on spectral slopes are typically in the range 0.1-—0.3, allow us only to focus on the more basic question, i.e.\ how it is possible to produce spectra steeper than $E^{-2}$ in SNRs.

Another effect which cannot be easily handled in this simplified approach is the dynamical role of the amplified magnetic field on the shock dynamics.
Pressure in the shape of magnetic turbulence may in fact dominate over ordinary thermal plasma pressure \emph{upstream}, leading to non trivial modifications of the Rankine-Hugoniot equations describing the jump conditions at the subshock. 
In the case of resonant Alfv\'en waves such a magnetic feedback has been shown to reduce the compressibility of the upstream plasma, providing a smoothening of the precursor (i.e.\ a reduction of the total compression ratio) and, as final consequence, steeper spectra of accelerated particles \cite{jumpl,veb07,jumpkin}. 
 
All of these arguments firmly asses the need for the inclusion of a detailed magnetic field treatment as a key ingredient in testing hadronic scenarios for the \gr~emission observed from SNRs. Nevertheless, the results outlined in this paper hold as a qualitative description and as a first-order estimates of many of the effects related to the observed magnetic field amplification ongoing in SNRs.   

\subsection{Molecular clouds}
The frequently observed correlation between \gr~bright regions around SNRs and MCs is indeed a strong hint in favor of a hadronic origin for such an emission.
There are in fact no evident reasons why the most plausible leptonic processes, namely relativistic bremsstrahlung and inverse-Compton scattering on the cosmic microwave background or on the Galactic infrared radiation, should correlate with the positions of MCs.
On the other hand, in a hadronic scenario a much more natural correlation is expected because of the huge amount of targets provided by environments where the density may easily be hundreds or thousands times larger than in the ISM.
 
For instance, in order to account for the observed luminosities from objects like IC443, W28N, W44, and W51C it is necessary to invoke local density as large as $n=10-100 cm^{-3}$. 
These sources are associated with MCs and are quite peculiar, in that they show a cut-off in the GeV band. 
They probably represent a somehow distinct class of objects (SNR-MC associations?) which cannot be described with the present simple model (they occupy a off-set positions in figure \ref{gevtev} as well).
Moreover, the luminosity in \gr s of these middle-age sources turns out to be as large as $10^{35}-10^{36}$erg s$^{-1}$, about one or two orders of magnitude above the most optimistic predictions in both figure \ref{power} and ref.~\cite{dav94}, which typically span the range $10^{33}-10^{34}$erg s$^{-1}$.
We cannot but conclude that, at least for these objects, the presence of huge repositories of molecular gas is fundamental in order to account for the large fluxes observed.

A detailed analysis of the properties of the \gr~emission expected from MCs is nevertheless beyond the goals of this paper, and hence the related phenomenology is not accounted for in the formalism above.
It is however necessary to spend a few words about their potential effects in the matter on debate.
 
Regarding the argument about the role of magnetic field amplification put forward in section \ref{sec:CR}, we can make the following considerations.
At the zeroth order, the role of a neutral component is to boost the luminosity, leaving only the slope unscathed. 
This is why it may be worth focusing more on the spectral slope rather than on the emission level.
At the next order, one could imagine the shock to slow down significantly because of the larger inertia of the swept-up material, but also the magnetic field to be suppressed because of a more efficient ion-neutral damping. 
Since these effects go in opposite directions in changing the Alfv\'enic Mach number, and may partially compensate themselves, it is hard to estimate \emph{a priori} the effects of a relevant neutral component on the particle spectrum. 

The actual role of MCs may however be twofold, depending on whether the forward shock has directly crashed into the cloud or not.
In the former case, the enhanced density of the cloud provides a huge amount of target nucleons, resulting in a magnified, direct probe of the CR content of the remnant, while in the latter case a MC may be illuminated only by \emph{escaping} CRs, whose spectrum is not strictly related to the one of particle being accelerated at the shock.
Some examples of multi-wavelength studies of the \gr~emission predicted from MCs illuminated by CRs escaping from close SNRs are put forward for instance in ref.~\cite{gab07,lke08,gac09,omy11}.

In any case, the NLDSA theory needed to predict the spectrum of the particles responsible for such an emission has to be completed with an additional ingredient not fully understood, yet.
In the first case, in fact, one should account for a shock propagating in a partially neutral environment, therefore including the effects of charge-exchange in the hydrodynamics and of ion-neutral damping in the magnetic field treatment; in the second case, instead, one should deal with the details of how particle escape from a SNR, a problem related with the evolution of the SNR confining power, and in turn, with the evolution of magnetic turbulence responsible for particle diffusion.
The transition from the diffusive regime inside the SNR and the one in the Galaxy, in fact, spans several order of magnitudes in the strength of the diffusion coefficient, $D(E)$; also the energy-dependence of the phenomenon is expected to vary from Bohm-like $D(E)\propto E$ to $D(E)\propto E^{\delta}$, with $\delta=0.3-0.6$, as inferred from secondary to primary ratios in diffuse Galactic CRs.
The exact details of such a poorly understood transition may dramatically change the predictions for the spectrum of escaping CRs impacting into a MC.

The problem of describing the escape of accelerated particles from a source is particularly important also because it is a fundamental piece of information needed to relate particle acceleration in SNRs to the diffuse spectrum of Galactic CRs.
In ref.~\cite{drury10,crspectrum} the reader can find a wider discussion providing some insights into why particle escape from SNRs still remains one of the most fundamental open question in the whole theory of the production of Galactic CRs.  

\subsection{SNR gamma-ray luminosity}
Present estimates of the time evolution of the \gr~luminosity of SNRs, however, must be taken with a grain of salt and always checked against non-linear calculations in which CR acceleration efficiency and maximum proton energy are calculated self-consistently.
A clear example of this is the rather unphysical growth of the predicted emission for very late stages: at a certain point the maximum energy of accelerated protons has to drop below a few GeV, implying a fading of the hadronic emission. 
In addition, it is still not clear whether the actual release of accelerated particles which have been advected downstream occurs only at the ``death'' of the SNR or whether it is a continuous process, occurring during the whole Sedov-Taylor stage \cite{crspectrum}.
If the latter were the case, it is possible for highest-energy particles to leave the remnant well before the beginning of the snowplow stage, hence severely limiting the TeV emission in old SNRs.
Nevertheless, it is worth remembering that the release of high-energy particles from downstream does not affect the global SNR evolution, since in our cases most of the energy in CRs responsible for the shell pressure is stored in mildly relativistic particles, spectra being steeper than $E^{-2}$.

The recipe for estimating the total \gr~emission adopted here and in ref.~\cite{dav94} neglects both the time evolution of $E_{max}$ and the details of the release of advected particles, therefore it is usually quite reliable for the GeV emission, but it has to be sometimes regarded only as a rule of thumb for multi-TeV photons.  

Another strictly related problem is the calculation of the probability of detecting a SNR as a \emph{PeVatron}, i.e.\ a source of photons as energetic as a few hundreds TeV produced by hadronic interactions of multi-PeV protons. 
Such an evidence would be, in fact, a clear-cut signature for acceleration of hadrons in SNRs, in addition up to the knee observed in the diffuse spectrum of Galactic CRs detected at Earth.
Again, the estimate of how many PeVatrons could be detected cannot fail to self-consistently account for CR acceleration efficiency, magnetic field amplification, evolution of $E_{max}$ and SNR luminosity in \gr s.
All of these quantities have to be calculated as a function of the circumstellar medium properties, and in particular have to be evaluated at the epoch of the transition between ejecta-dominated and Sedov-Taylor stages. 
In fact, a SNR might be expected to work as a PeVatron only during this peculiar evolutionary stage \cite{escape,crspectrum}.   

Another variable that may enter the history of the \gr~emission from SNRs is the abundance of nuclei heavier than Hydrogen, both in the circumstellar medium and in the accelerated particles. 
Taking into account the interstellar abundances of all the chemical elements, in fact, increases the rate of nuclear collisions by a factor 1.5--2 \cite{mori09}, while the abundances of heavy nuclei in accelerated particles, inferred by the CR chemical composition measured at Earth, typically boost the total secondary emission by another factor 2--3, as shown in \cite{nuclei}.  
Very generally, the chemical composition of the wind and of the bubble may be quite different from the ISM one, and such a difference might have observational consequences on the level of the expected \gr~ emission and, in turn, in the statistical considerations above.

\section{Conclusions}\label{sec:conclusion}
In this work we tried to understand the impact of very recent observations upon the so-called hadronic scenario for the \gr~emission from SNRs.
When energetic photons result from the decay of neutral pions produced in nuclear interactions between relativistic hadrons and background plasma, their spectrum \emph{unequivocally} maps the one of accelerated particles, providing an insight into the instantaneous content of a SNR in terms of cosmic rays.

In table \ref{tab:SNR} we collected the most, as far as we know, updated list of \gr-bright SNRs and of \gr-sources potentially associated with SNRs, showing that basically all of their energy spectra are invariably steeper than $E^{-2}$.
As outlined in section \ref{sec:intro}, such steep spectra are apparently at odds with standard expectations of NLDSA theories at SNR shocks, which predict the spectra of high-energy particles to be flatter than $E^{-2}$ when the acceleration is efficient.

In section \ref{sec:CR} we illustrated a possible way to reconcile NLDSA predictions with observations, namely by accounting for the generation of magnetic turbulence induced by the same accelerated particles. 
In fact, when resonant streaming instability is as effective as to reproduce the large magnetic field inferred in young SNRs (a few hundreds $\mu$G), the standard picture of the particle-wave scattering is, very likely, no longer suitable for describing CR transport.
In particular, the velocity of the magnetic perturbations particles scatter against may be not negligible with respect to the fluid velocity so that, when this occurs, a sizable modification of the average compression ratios felt by diffusing particles is expected, eventually affecting the spectrum of accelerated particles.
Such an effect has already been put forward before (see e.g.\ \cite{bell78a}), but its quantitative implications, especially under the light of the more recent discovery of effective magnetic field amplification in young SNRs, have not been deeply investigated yet.
This key ingredient may actually reverse the role of the CR acceleration efficiency in the determination of the spectral slope, predicting that the larger the CR pressure, the more efficient the amplification of the magnetic field and, in turn, the steeper the spectra of the accelerated particles.  
 
It is also worth stressing that the need for a physical mechanism producing steep spectra is required when explaining the detected emission as either due to relativistic bremsstrahlung or to pion decay, since in both cases the produced \gr~spectrum is parallel to the energy distribution of the parent particles. 
Strictly speaking, only inverse-Compton emission would not require to accelerate particles with a spectrum steeper than $E^{-2}$ but, on the other hand, such an emission mechanism tends to produce rather flat photon spectra, at odds with broadband observations listed in table \ref{tab:SNR} (see also comments in section \ref{sec:gevtev}). 
The spirit of the calculations put forward in section \ref{sec:CR} may thus apply not only in hadronic scenarios, but even in several other situations where the emission is thought be predominantly leptonic and due to relativistic bremsstrahlung, like in Cas A.

We showed in section \ref{sec:comparison} that a CR acceleration efficiency between 10 and 20\% may easily account for the spectra observed in most of the detected SNRs (figure \ref{gevtev}). 
In our calculations, the expected \gr~spectrum is calculated during the different evolutionary stages of a  remnant produced by a core-collapse SN. 
This means that we also included the non trivial circumstellar profile induced by stellar winds launched during the red-giant and Wolf-Rayet stages of the SN progenitor.
In addition, we worked out the SNR luminosity in the GeV (namely, above 100 MeV) and in the TeV band and found that it has a minimum when the remnant is between 5 and 10 thousands years old, corresponding to the propagation of the forward shock into the hot and rarefied bubble excavated by the Wolf-Rayet wind, a feature that seems to be recovered also in figure \ref{gevtev}. 

Finally, in section \ref{sec:cav} we discussed strong points and limitations of the present approach, highlighting open theoretical problems, as outlining a consistent description of the non-linear interplay between CRs and magnetic field, but also observational issues, as the observed correlation between strong \gr~emissivity and MCs.

Our findings, worked out in the context of a rather phenomenological scenario, have however to be checked against fully non-linear calculations able to self-consistently account for shock dynamics, particle acceleration and magnetic field amplification, but they indeed represent a preliminary, necessary step for understanding hadronic emission from SNRs.
Under the light of this revised theoretical framework for the hadronic emission, we finally discussed what future observations achievable with the present (and next) generation of \gr~telescopes may tell us, not only regarding non-thermal properties of SNRs, but also regarding the origin of Galactic CRs.

\acknowledgments
The author is grateful to P. Blasi, E. Amato, G. Morlino and R. Bandiera for their support and for having read a preliminary version of this paper. The author also thanks an anonymous referee for her/his valuable comments. 
This work was partially supported by ASI through contract ASI-INAF I/088/06/0. 

\bibliographystyle{JHEP}
\bibliography{bibgamma}

\providecommand{\href}[2]{#2}\begingroup\raggedright\begin{thebibliography}{10%
0}

\bibitem{baade-zwicky34}
W.~{Baade} and F.~{Zwicky}, {\it {Cosmic Rays from Super-novae}},  {\em
  Proceedings of the National Academy of Science} {\bf 20} (1934) 259--263.

\bibitem{dav94}
L.~{O'C. Drury}, F.~{Aharonian}, and H.~J. {V{\"o}lk}, {\it {The gamma-ray
  visibility of supernova remnants. A test of cosmic ray origin}},  {\em \aap}
  {\bf 287} (July, 1994) 959--971,
  [\href{http://xxx.lanl.gov/abs/astro-ph/9305037}{{\tt astro-ph/9305037}}].

\bibitem{enomoto+02}
R.~{Enomoto et al.}, {\it {The acceleration of cosmic-ray protons in the
  supernova remnant RX J1713.7-3946}},  {\em Nature} {\bf 416} (Apr., 2002)
  823--826.

\bibitem{Fermi1713}
A.~A. {Abdo} and {Fermi LAT Collaboration}, {\it {Observations of the young
  supernova remnant RX J1713.7-3946 with the Fermi Large Area Telescope}},
  {\em ArXiv e-prints} (Mar., 2011)
  [\href{http://xxx.lanl.gov/abs/1103.5727}{{\tt arXiv:1103.5727}}].

\bibitem{bv06}
E.~G. {Berezhko} and H.~J. {V{\"o}lk}, {\it {Theory of cosmic ray production in
  the supernova remnant RX J1713.7-3946}},  {\em \aap} {\bf 451} (June, 2006)
  981--990, [\href{http://xxx.lanl.gov/abs/astro-ph/0602177}{{\tt
  astro-ph/0602177}}].

\bibitem{gio+09}
G.~{Morlino}, E.~{Amato}, and P.~{Blasi}, {\it {Gamma-ray emission from SNR RX
  J1713.7-3946 and the origin of galactic cosmic rays}},  {\em MNRAS} {\bf 392}
  (Jan., 2009) 240--250, [\href{http://xxx.lanl.gov/abs/0810.0094}{{\tt
  arXiv:0810.0094}}].

\bibitem{za10}
V.~N. {Zirakashvili} and F.~A. {Aharonian}, {\it {Nonthermal Radiation of Young
  Supernova Remnants: The Case of RX J1713.7-3946}},  {\em ApJ} {\bf 708}
  (Jan., 2010) 965--980, [\href{http://xxx.lanl.gov/abs/0909.2285}{{\tt
  arXiv:0909.2285}}].

\bibitem{ellison+10}
D.~C. {Ellison}, D.~J. {Patnaude}, P.~{Slane}, and J.~{Raymond}, {\it
  {Efficient Cosmic Ray Acceleration, Hydrodynamics, and Self-Consistent
  Thermal X-Ray Emission Applied to Supernova Remnant RX J1713.7-3946}},  {\em
  \apj} {\bf 712} (Mar., 2010) 287--293,
  [\href{http://xxx.lanl.gov/abs/1001.1932}{{\tt arXiv:1001.1932}}].

\bibitem{fang+09}
J.~{Fang}, L.~{Zhang}, J.~F. {Zhang}, Y.~Y. {Tang}, and H.~{Yu}, {\it {The
  multi-band non-thermal emission from the supernova remnant RX J1713.7-3946}},
   {\em \mnras} {\bf 392} (Jan., 2009) 925--929,
  [\href{http://xxx.lanl.gov/abs/0810.4097}{{\tt arXiv:0810.4097}}].

\bibitem{casanova+10}
S.~{Casanova}, D.~I. {Jones}, F.~A. {Aharonian}, Y.~{Fukui}, S.~{Gabici},
  A.~{Kawamura}, T.~{Onishi}, G.~{Rowell}, H.~{Sano}, K.~{Torii}, and
  H.~{Yamamoto}, {\it {Modeling the Gamma-Ray Emission Produced by Runaway
  Cosmic Rays in the Environment of RX J1713.7$-$3946}},  {\em \pasj} {\bf 62}
  (Oct., 2010) 1127--1134, [\href{http://xxx.lanl.gov/abs/1003.0379}{{\tt
  arXiv:1003.0379}}].

\bibitem{plaga08}
R.~{Plaga}, {\it {Arguments against a dominantly hadronic origin of the VHE
  radiation from the supernova remnant RX J1713-3946}},  {\em New Astronomy}
  {\bf 13} (Feb., 2008) 73--76, [\href{http://xxx.lanl.gov/abs/0711.4046}{{\tt
  arXiv:0711.4046}}].

\bibitem{yuan+10}
Q.~{Yuan}, S.~{Liu}, Z.~{Fan}, X.~{Bi}, and C.~L. {Fryer}, {\it {Modeling the
  Multi-Wavelength Emission of Shell-Type Supernova Remnant RX J1713.7-3946}},
  {\em ArXiv e-prints} (Oct., 2010)
  [\href{http://xxx.lanl.gov/abs/1011.0145}{{\tt arXiv:1011.0145}}].

\bibitem{Fang+11}
J.~{Fang}, Y.~{Tang}, and L.~{Zhang}, {\it {Multiband Nonthermal Emission from
  Three TeV Shell-type Supernova Remnants}},  {\em \apj} {\bf 731} (Apr., 2011)
  32--+.

\bibitem{jiang+10}
B.~{Jiang}, Y.~{Chen}, J.~{Wang}, Y.~{Su}, X.~{Zhou}, S.~{Safi-Harb}, and
  T.~{DeLaney}, {\it {Cavity of Molecular Gas Associated with Supernova Remnant
  3C 397}},  {\em \apj} {\bf 712} (Apr., 2010) 1147--1156,
  [\href{http://xxx.lanl.gov/abs/1001.2204}{{\tt arXiv:1001.2204}}].

\bibitem{axford+77}
W.~I. {Axford}, E.~{Leer}, and G.~{Skadron}, {\it {Acceleration of Cosmic Rays
  at Shock Fronts (Abstract)}},  in {\em \emph{Acceleration of Cosmic Rays at
  Shock Fronts}}, vol.~2 of {\em International Cosmic Ray Conference},
  pp.~273--+, 1977.

\bibitem{krymskii77}
G.~F. {Krymskii}, {\it {A regular mechanism for the acceleration of charged
  particles on the front of a shock wave}},  {\em Akademiia Nauk SSSR Doklady}
  {\bf 234} (June, 1977) 1306--1308.

\bibitem{ALS78}
W.~I. {Axford}, E.~{Leer}, and G.~{Skadron}, {\it {The acceleration of cosmic
  rays by shock waves}},  in {\em \emph{The acceleration of cosmic rays by
  shock waves}}, vol.~11 of {\em International Cosmic Ray Conference},
  pp.~132--137, 1978.

\bibitem{blandford-ostriker78}
R.~D. {Blandford} and J.~P. {Ostriker}, {\it {Particle acceleration by
  astrophysical shocks}},  {\em ApJL} {\bf 221} (Apr., 1978) L29--L32.

\bibitem{bell78a}
A.~R. {Bell}, {\it {The acceleration of cosmic rays in shock fronts. I}},  {\em
  MNRAS} {\bf 182} (Jan., 1978) 147--156.

\bibitem{bell78b}
A.~R. {Bell}, {\it {The acceleration of cosmic rays in shock fronts. II}},
  {\em MNRAS} {\bf 182} (Feb., 1978) 443--455.

\bibitem{drury-volk81a}
L.~{O'C. Drury} and H.~J. {V{\"o}lk}, {\it {Hydromagnetic shock structure in
  the presence of cosmic rays}},  {\em Ap. J.} {\bf 248} (Aug., 1981) 344--351.

\bibitem{drury-volk81b}
L.~{O'C. Drury} and H.~J. {V{\"o}lk}, {\it {Shock structure including cosmic
  ray acceleration}},  in {\em \emph{Origin of Cosmic Rays: Shock structure
  including cosmic ray acceleration}}, vol.~94 of {\em IAU Symposium},
  pp.~363--+, 1981.

\bibitem{drury83}
L.~{O'C. Drury}, {\it {An introduction to the theory of diffusive shock
  acceleration of energetic particles in tenuous plasmas}},  {\em Reports of
  Progress in Physics} {\bf 46} (Aug., 1983) 973--1027.

\bibitem{blandford-eichler87}
R.~{Blandford} and D.~{Eichler}, {\it {Particle acceleration at astrophysical
  shocks: A theory of cosmic ray origin}},  {\em \physrep} {\bf 154} (Oct.,
  1987) 1--75.

\bibitem{jones-ellison91}
F.~C. {Jones} and D.~C. {Ellison}, {\it {The plasma physics of shock
  acceleration}},  {\em Space Science Reviews} {\bf 58} (1991) 259--346.

\bibitem{malkov-drury01}
M.~A. {Malkov} and L.~{O'C. Drury}, {\it {Nonlinear theory of diffusive
  acceleration of particles by shock waves }},  {\em Reports of Progress in
  Physics} {\bf 64} (Apr., 2001) 429--481.

\bibitem{comparison}
D.~{Caprioli}, H.~{Kang}, A.~E. {Vladimirov}, and T.~W. {Jones}, {\it
  {Comparison of different methods for non-linear diffusive shock
  acceleration}},  {\em \mnras} {\bf 407} (Sept., 2010) 1773--1783,
  [\href{http://xxx.lanl.gov/abs/1005.2127}{{\tt arXiv:1005.2127}}].

\bibitem{chevalier83}
R.~A. {Chevalier}, {\it {Blast waves with cosmic-ray pressure}},  {\em \apj}
  {\bf 272} (Sept., 1983) 765--772.

\bibitem{blondin-ellison01}
J.~M. {Blondin} and D.~C. {Ellison}, {\it {Rayleigh-Taylor Instabilities in
  Young Supernova Remnants Undergoing Efficient Particle Acceleration}},  {\em
  \apj} {\bf 560} (Oct., 2001) 244--253,
  [\href{http://xxx.lanl.gov/abs/astro-ph/0104024}{{\tt astro-ph/0104024}}].

\bibitem{ab06}
E.~{Amato} and P.~{Blasi}, {\it {Non-linear particle acceleration at
  non-relativistic shock waves in the presence of self-generated turbulence}},
  {\em MNRAS} {\bf 371} (Sept., 2006) 1251--1258,
  [\href{http://xxx.lanl.gov/abs/astro-ph/0606592}{{\tt astro-ph/0606592}}].

\bibitem{W28Fermi}
A.~A. {Abdo et al.}, {\it {Fermi Large Area Telescope Observations of the
  Supernova Remnant W28 (G6.4-0.1)}},  {\em \apj} {\bf 718} (July, 2010)
  348--356.

\bibitem{W28AGILE}
A.~{Giuliani et al.}, {\it {AGILE detection of GeV {$\gamma$}-ray emission from
  the SNR W28}},  {\em \aap} {\bf 516} (June, 2010) L11+,
  [\href{http://xxx.lanl.gov/abs/1005.0784}{{\tt arXiv:1005.0784}}].

\bibitem{W28HESS}
F.~{Aharonian et al.}, {\it {Discovery of very high energy gamma-ray emission
  coincident with molecular clouds in the W 28 (G6.4-0.1) field}},  {\em \aap}
  {\bf 481} (Apr., 2008) 401--410,
  [\href{http://xxx.lanl.gov/abs/0801.3555}{{\tt arXiv:0801.3555}}].

\bibitem{cg08}
J.~{Casandjian} and I.~A. {Grenier}, {\it {A revised catalogue of EGRET
  {$\gamma$}-ray sources}},  {\em \aap} {\bf 489} (Oct., 2008) 849--883,
  [\href{http://xxx.lanl.gov/abs/0806.0113}{{\tt arXiv:0806.0113}}].

\bibitem{cs10}
D.~{Castro} and P.~{Slane}, {\it {Fermi Large Area Telescope Observations of
  Supernova Remnants Interacting with Molecular Clouds}},  {\em \apj} {\bf 717}
  (July, 2010) 372--378, [\href{http://xxx.lanl.gov/abs/1002.2738}{{\tt
  arXiv:1002.2738}}].

\bibitem{aha+06}
F.~{Aharonian et al.}, {\it {The H.E.S.S. Survey of the Inner Galaxy in Very
  High Energy Gamma Rays}},  {\em \apj} {\bf 636} (Jan., 2006) 777--797,
  [\href{http://xxx.lanl.gov/abs/astro-ph/0510397}{{\tt astro-ph/0510397}}].

\bibitem{W30CANGAROO}
Y.~{Higashi et al.}, {\it {Observation of Very High Energy Gamma Rays from HESS
  J1804-216 with CANGAROO-III Telescopes}},  {\em \apj} {\bf 683} (Aug., 2008)
  957--966, [\href{http://xxx.lanl.gov/abs/0805.0708}{{\tt arXiv:0805.0708}}].

\bibitem{0FGL}
A.~A. {Abdo et al.}, {\it {Fermi/Large Area Telescope Bright Gamma-Ray Source
  List}},  {\em \apjs} {\bf 183} (July, 2009) 46--66,
  [\href{http://xxx.lanl.gov/abs/0902.1340}{{\tt arXiv:0902.1340}}].

\bibitem{W44Fermi}
A.~A. {Abdo et al.}, {\it {Gamma-Ray Emission from the Shell of Supernova
  Remnant W44 Revealed by the Fermi LAT}},  {\em Science} {\bf 327} (Feb.,
  2010) 1103--.

\bibitem{W49BFermi}
A.~A. {Abdo et al.}, {\it {Fermi-LAT Study of Gamma-ray Emission in the
  Direction of Supernova Remnant W49B}},  {\em \apj} {\bf 722} (Oct., 2010)
  1303--1311.

\bibitem{W51CFermi}
A.~A. {Abdo et al.}, {\it {Fermi LAT Discovery of Extended Gamma-Ray Emission
  in the Direction of Supernova Remnant W51C}},  {\em \apjl} {\bf 706} (Nov.,
  2009) L1--L6, [\href{http://xxx.lanl.gov/abs/0910.0908}{{\tt
  arXiv:0910.0908}}].

\bibitem{W51CHESS}
A.~{Fiasson}, K.~{Kosack}, J.~{Skilton}, Y.~{Gallant}, J.~{Hinton}, and
  G.~{P{\"u}lhofer}, {\it {Probing cosmic ray acceleration through molecular
  clouds in the vicinity of supernova remnant with H.E.S.S.}},  in {\em
  American Institute of Physics Conference Series} ({F.~A.~Aharonian,
  W.~Hofmann, \& F.~Rieger}, ed.), vol.~1085 of {\em American Institute of
  Physics Conference Series}, pp.~361--363, Dec., 2008.

\bibitem{MILAGRO09}
A.~A. {Abdo et al.}, {\it {Milagro Observations of Multi-TeV Emission from
  Galactic Sources in the Fermi Bright Source List \emph{[Erratum: }{\bf 703}
  \apjl~\emph{(2009) L185]}}},  {\em \apjl} {\bf 700} (Aug., 2009) L127--L131,
  [\href{http://xxx.lanl.gov/abs/0904.1018}{{\tt arXiv:0904.1018}}].

\bibitem{G106VERITAS}
V.~A. {Acciari et al.}, {\it {Detection of Extended VHE Gamma Ray Emission from
  G106.3+2.7 with Veritas}},  {\em \apjl} {\bf 703} (Sept., 2009) L6--L9,
  [\href{http://xxx.lanl.gov/abs/0911.4695}{{\tt arXiv:0911.4695}}].

\bibitem{CasAFermi}
A.~A. {Abdo et al.}, {\it {Fermi-Lat Discovery of GeV Gamma-Ray Emission from
  the Young Supernova Remnant Cassiopeia A}},  {\em \apjl} {\bf 710} (Feb.,
  2010) L92--L97, [\href{http://xxx.lanl.gov/abs/1001.1419}{{\tt
  arXiv:1001.1419}}].

\bibitem{CasAVERITAS}
V.~A. {Acciari et al.}, {\it {Observations of the Shell-type Supernova Remnant
  Cassiopeia A at TeV Energies with VERITAS}},  {\em \apj} {\bf 714} (May,
  2010) 163--169, [\href{http://xxx.lanl.gov/abs/1002.2974}{{\tt
  arXiv:1002.2974}}].

\bibitem{CasAMAGIC}
J.~{Albert et al.}, {\it {Observation of VHE {$\gamma$}-rays from Cassiopeia A
  with the MAGIC telescope}},  {\em \aap} {\bf 474} (Nov., 2007) 937--940,
  [\href{http://xxx.lanl.gov/abs/0706.4065}{{\tt arXiv:0706.4065}}].

\bibitem{TychoFermi}
F.~{Giordano for the \emph{Fermi} collaboration}, {\it {GeV Observations of the
  remnants of historical supernovae}},  {\em \emph{Poster at the 25$^{th}$
  Texas Symposium on Relativistic Astrophysics, Heidelberg (Germany); Session
  {\bf P5}: Galactic and Extragalactic Cosmic Rays}} (Dec., 2010).

\bibitem{TychoVER}
V.~A. {Acciari et al.}, {\it {Discovery of TeV Gamma Ray Emission from Tycho's
  Supernova Remnant}},  {\em ArXiv e-prints} (Feb., 2011)
  [\href{http://xxx.lanl.gov/abs/1102.3871}{{\tt arXiv:1102.3871}}].

\bibitem{IC443Fermi}
A.~A. {Abdo et al.}, {\it {Observation of Supernova Remnant IC 443 with the
  Fermi Large Area Telescope}},  {\em \apj} {\bf 712} (Mar., 2010) 459--468,
  [\href{http://xxx.lanl.gov/abs/1002.2198}{{\tt arXiv:1002.2198}}].

\bibitem{IC443AGILE}
M.~{Tavani et al.}, {\it {Direct Evidence for Hadronic Cosmic-Ray Acceleration
  in the Supernova Remnant IC 443}},  {\em \apjl} {\bf 710} (Feb., 2010)
  L151--L155, [\href{http://xxx.lanl.gov/abs/1001.5150}{{\tt
  arXiv:1001.5150}}].

\bibitem{IC443MAGIC}
J.~{Albert et al.}, {\it {Discovery of Very High Energy Gamma Radiation from IC
  443 with the MAGIC Telescope}},  {\em \apjl} {\bf 664} (Aug., 2007) L87--L90,
  [\href{http://xxx.lanl.gov/abs/0705.3119}{{\tt arXiv:0705.3119}}].

\bibitem{IC443VERITAS}
V.~A. {Acciari et al.}, {\it {Observation of Extended Very High Energy Emission
  from the Supernova Remnant IC 443 with VERITAS}},  {\em \apjl} {\bf 698}
  (June, 2009) L133--L137, [\href{http://xxx.lanl.gov/abs/0905.3291}{{\tt
  arXiv:0905.3291}}].

\bibitem{wang-scoville92}
Z.~{Wang} and N.~Z. {Scoville}, {\it {Strongly shocked interstellar gas in IC
  443. I - High-resolution molecular observations}},  {\em \apj} {\bf 386}
  (Feb., 1992) 158--169.

\bibitem{MonocerosHESS}
F.~A. {Aharonian et al.}, {\it {Discovery of a point-like very-high-energy
  {$\gamma$}-ray source in Monoceros}},  {\em \aap} {\bf 469} (July, 2007)
  L1--L4, [\href{http://xxx.lanl.gov/abs/0704.0171}{{\tt arXiv:0704.0171}}].

\bibitem{VelaCANGAROO}
H.~{Katagiri et al.}, {\it {Detection of Gamma Rays around 1 TeV from RX
  J0852.0-4622 by CANGAROO-II}},  {\em \apjl} {\bf 619} (Feb., 2005)
  L163--L166, [\href{http://xxx.lanl.gov/abs/astro-ph/0412623}{{\tt
  astro-ph/0412623}}].

\bibitem{VelaHESS}
F.~{Aharonian et al.}, {\it {H.E.S.S. Observations of the Supernova Remnant RX
  J0852.0-4622: Shell-Type Morphology and Spectrum of a Widely Extended Very
  High Energy Gamma-Ray Source}},  {\em \apj} {\bf 661} (May, 2007) 236--249,
  [\href{http://xxx.lanl.gov/abs/astro-ph/0612495}{{\tt astro-ph/0612495}}].

\bibitem{RCW86HESS}
F.~{Aharonian et al.}, {\it {Discovery of Gamma-Ray Emission From the
  Shell-Type Supernova Remnant RCW 86 With Hess}},  {\em \apj} {\bf 692} (Feb.,
  2009) 1500--1505, [\href{http://xxx.lanl.gov/abs/0810.2689}{{\tt
  arXiv:0810.2689}}].

\bibitem{SN1006HESS}
F.~{Acero et al.}, {\it {First detection of VHE {$\gamma$}-rays from SN 1006 by
  HESS}},  {\em \aap} {\bf 516} (June, 2010) A62+,
  [\href{http://xxx.lanl.gov/abs/1004.2124}{{\tt arXiv:1004.2124}}].

\bibitem{RXJFermi}
M.~{Lemoine-Goumard for the \emph{Fermi} collaboration}, {\it {SNR with
  Fermi-LAT after 1.5 year of Observations}},  {\em \emph{SNR-PWN Workshop in
  Montpellier (France)}} (May, 2010).

\bibitem{RXJ1713HESS}
F.~{Aharonian et al.}, {\it {A detailed spectral and morphological study of the
  gamma-ray supernova remnant RX J1713.7-3946 with HESS}},  {\em A\&A} {\bf
  449} (Apr., 2006) 223--242,
  [\href{http://xxx.lanl.gov/abs/astro-ph/0511678}{{\tt astro-ph/0511678}}].

\bibitem{uchiyama+07}
Y.~{Uchiyama}, F.~A. {Aharonian}, T.~{Tanaka}, T.~{Takahashi}, and Y.~{Maeda},
  {\it {Extremely fast acceleration of cosmic rays in a supernova remnant}},
  {\em Nature} {\bf 449} (Oct., 2007) 576--578.

\bibitem{RXJCANGAROO}
H.~{Muraishi et al.}, {\it {Evidence for TeV gamma-ray emission from the shell
  type SNR RX J1713.7-3946}},  {\em \aap} {\bf 354} (Feb., 2000) L57--L61,
  [\href{http://xxx.lanl.gov/abs/astro-ph/0001047}{{\tt astro-ph/0001047}}].

\bibitem{CTB37HESS}
F.~{Aharonian et al.}, {\it {Discovery of a VHE gamma-ray source coincident
  with the supernova remnant CTB 37A}},  {\em \aap} {\bf 490} (Nov., 2008)
  685--693, [\href{http://xxx.lanl.gov/abs/0803.0702}{{\tt arXiv:0803.0702}}].

\bibitem{Tian+10}
W.~W. {Tian}, Z.~{Li}, D.~A. {Leahy}, J.~{Yang}, X.~J. {Yang}, R.~{Yamazaki},
  and D.~{Lu}, {\it {X-Ray Emission from HESS J1731-347/SNR G353.6-0.7 and
  Central Compact Source XMMS J173203-344518}},  {\em \apj} {\bf 712} (Apr.,
  2010) 790--796, [\href{http://xxx.lanl.gov/abs/0907.1684}{{\tt
  arXiv:0907.1684}}].

\bibitem{HESS1731}
F.~{Aharonian et al.}, {\it {HESS very-high-energy gamma-ray sources without
  identified counterparts}},  {\em \aap} {\bf 477} (Jan., 2008) 353--363,
  [\href{http://xxx.lanl.gov/abs/0712.1173}{{\tt arXiv:0712.1173}}].

\bibitem{fm05}
M.~{Fatuzzo} and F.~{Melia}, {\it {Primary versus Secondary Leptons in the
  EGRET Supernova Remnants}},  {\em \apj} {\bf 630} (Sept., 2005) 321--331,
  [\href{http://xxx.lanl.gov/abs/astro-ph/0412468}{{\tt astro-ph/0412468}}].

\bibitem{dja+08}
A.~{Djannati-Ata{\u i}}, O.~C. {de Jager}, R.~{Terrier}, and {et al.}, {\it
  {New Companions for the lonely Crab? VHE emission from young pulsar wind
  nebulae revealed by H.E.S.S.}},  in {\em International Cosmic Ray
  Conference}, vol.~2 of {\em International Cosmic Ray Conference},
  pp.~823--826, 2008.

\bibitem{W41MAGIC}
J.~{Albert et al.}, {\it {Observation of VHE Gamma Radiation from HESS
  J1834-087/W41 with the MAGIC Telescope}},  {\em \apjl} {\bf 643} (May, 2006)
  L53--L56, [\href{http://xxx.lanl.gov/abs/astro-ph/0604197}{{\tt
  astro-ph/0604197}}].

\bibitem{Paron+11}
S.~{Paron}, E.~{Giacani}, M.~{Rubio}, and G.~{Dubner}, {\it {Study of the
  molecular clump associated with the high-energy source HESS J1858+020}},
  {\em ArXiv e-prints} (Mar., 2011)
  [\href{http://xxx.lanl.gov/abs/1103.5742}{{\tt arXiv:1103.5742}}].

\bibitem{G405}
F.~{Aharonian et al}, {\it {Detection of very high energy radiation from HESS
  J1908+063 confirms the Milagro unidentified source MGRO J1908+06}},  {\em
  \aap} {\bf 499} (June, 2009) 723--728,
  [\href{http://xxx.lanl.gov/abs/0904.3409}{{\tt arXiv:0904.3409}}].

\bibitem{wak+10}
S.~{Wakely} and {VERITAS Collaboration}, {\it {Recent Veritas Observations of
  PWN and SNR Systems}},  in {\em Bulletin of the American Astronomical
  Society}, vol.~42 of {\em Bulletin of the American Astronomical Society},
  pp.~688--+, Feb., 2010.

\bibitem{tl06}
W.~W. {Tian} and D.~A. {Leahy}, {\it {The radio SNR G65.1+0.6 and its
  associated pulsar J1957+2831}},  {\em \aap} {\bf 455} (Sept., 2006)
  1053--1058, [\href{http://xxx.lanl.gov/abs/astro-ph/0603102}{{\tt
  astro-ph/0603102}}].

\bibitem{weinstein09}
A.~{Weinstein} and {for the VERITAS Collaboration}, {\it {The VERITAS Survey of
  the Cygnus Region of the Galactic Plane}},  {\em ArXiv e-prints} (Dec., 2009)
  [\href{http://xxx.lanl.gov/abs/0912.4492}{{\tt arXiv:0912.4492}}].

\bibitem{HESSJ1507}
F.~{Acero et al}, {\it {Discovery and follow-up studies of the extended,
  off-plane, VHE gamma-ray source HESS J1507-622}},  {\em \aap} {\bf 525}
  (Jan., 2011) A45+, [\href{http://xxx.lanl.gov/abs/1010.4907}{{\tt
  arXiv:1010.4907}}].

\bibitem{tibolla09}
O.~{Tibolla}, R.~C.~G. {Chaves}, W.~{Domainko}, O.~{de Jager}, S.~{Kaufmann},
  S.~{Wagner}, N.~{Komin}, K.~{Kosack}, A.~{Fiasson}, M.~{Renaud}, and {for the
  H.~E.~S.~S.~Collaboration}, {\it {New unidentified Galactic H.E.S.S.
  sources}},  {\em ArXiv e-prints} (Dec., 2009)
  [\href{http://xxx.lanl.gov/abs/0912.3811}{{\tt arXiv:0912.3811}}].

\bibitem{HESS1745}
F.~{Aharonian et al.}, {\it {Exploring a SNR/molecular cloud association within
  HESS J1745-303}},  {\em \aap} {\bf 483} (May, 2008) 509--517,
  [\href{http://xxx.lanl.gov/abs/0803.2844}{{\tt arXiv:0803.2844}}].

\bibitem{Tian11}
W.~W. {Tian} and D.~A. {Leahy}, {\it {Tycho SN 1572: A Naked Ia Supernova
  Remnant Without an Associated Ambient Molecular Cloud}},  {\em \apjl} {\bf
  729} (Mar., 2011) L15+, [\href{http://xxx.lanl.gov/abs/1012.5673}{{\tt
  arXiv:1012.5673}}].

\bibitem{SNRdata1}
O.~H. {Guseinov}, A.~{Ankay}, and S.~O. {Tagieva}, {\it {Observational data on
  galactic supernova remnants: the supernova remnants within l=0-90 degrees}},
  {\em Serbian Astronomical Journal} {\bf 167} (Dec., 2003) 93--+.

\bibitem{SNRdata2}
O.~H. {Guseinov}, A.~{Ankay}, and S.~O. {Tagieva}, {\it {Observational data on
  galactic supernova remnants: the supernova remnants within l=90-270
  degrees}},  {\em Serbian Astronomical Journal} {\bf 168} (Feb., 2004) 55--+.

\bibitem{SNRdata3}
O.~H. {Guseinov}, A.~{Ankay}, and S.~O. {Tagieva}, {\it {Observational data on
  galactic supernova remnants: the supernova remnants within l=270-360
  degrees}},  {\em Serbian Astronomical Journal} {\bf 169} (Mar., 2004) 65--+.

\bibitem{Bamba+05}
A.~{Bamba}, R.~{Yamazaki}, T.~{Yoshida}, T.~{Terasawa}, and K.~{Koyama}, {\it
  {A Spatial and Spectral Study of Nonthermal Filaments in Historical Supernova
  Remnants: Observational Results with Chandra}},  {\em \apj} {\bf 621} (Mar.,
  2005) 793--802, [\href{http://xxx.lanl.gov/abs/astro-ph/0411326}{{\tt
  astro-ph/0411326}}].

\bibitem{V+05}
H.~J. {V{\"o}lk}, E.~G. {Berezhko}, and L.~T. {Ksenofontov}, {\it {Magnetic
  field amplification in Tycho and other shell-type supernova remnants}},  {\em
  A\&A} {\bf 433} (Apr., 2005) 229--240,
  [\href{http://xxx.lanl.gov/abs/astro-ph/0409453}{{\tt astro-ph/0409453}}].

\bibitem{P+06}
E.~{Parizot}, A.~{Marcowith}, J.~{Ballet}, and Y.~A. {Gallant}, {\it
  {Observational constraints on energetic particle diffusion in young
  supernovae remnants: amplified magnetic field and maximum energy}},  {\em
  A\&A} {\bf 453} (July, 2006) 387--395,
  [\href{http://xxx.lanl.gov/abs/astro-ph/0603723}{{\tt astro-ph/0603723}}].

\bibitem{eriksen+11}
K.~A. {Eriksen}, J.~P. {Hughes}, C.~{Badenes}, R.~{Fesen}, P.~{Ghavamian},
  D.~{Moffett}, P.~P. {Plucinksy}, C.~E. {Rakowski}, E.~M. {Reynoso}, and
  P.~{Slane}, {\it {Evidence for Particle Acceleration to the Knee of the
  Cosmic Ray Spectrum in Tycho's Supernova Remnant}},  {\em \apjl} {\bf 728}
  (Feb., 2011) L28+, [\href{http://xxx.lanl.gov/abs/1101.1454}{{\tt
  arXiv:1101.1454}}].

\bibitem{skilling75a}
J.~{Skilling}, {\it {Cosmic ray streaming. I - Effect of Alfven waves on
  particles}},  {\em MNRAS} {\bf 172} (Sept., {1975a}) 557--566.

\bibitem{skilling75b}
J.~{Skilling}, {\it {Cosmic ray streaming. II - Effect of particles on Alfven
  waves}},  {\em MNRAS} {\bf 173} (Nov., {1975b}) 245--254.

\bibitem{skilling75c}
J.~{Skilling}, {\it {Cosmic ray streaming. III - Self-consistent solutions}},
  {\em MNRAS} {\bf 173} (Nov., {1975c}) 255--269.

\bibitem{bell04}
A.~R. {Bell}, {\it {Turbulent amplification of magnetic field and diffusive
  shock acceleration of cosmic rays}},  {\em MNRAS} {\bf 353} (Sept., 2004)
  550--558.

\bibitem{bell05}
A.~R. {Bell}, {\it {The interaction of cosmic rays and magnetized plasma}},
  {\em MNRAS} {\bf 358} (Mar., 2005) 181--187.

\bibitem{ab08}
E.~{Amato} and P.~{Blasi}, {\it {A kinetic approach to cosmic ray induced
  streaming instability at supernova shocks}},  {\em astro-ph/0806.1223} (June,
  2008) [\href{http://xxx.lanl.gov/abs/0806.1223}{{\tt arXiv:0806.1223}}].

\bibitem{zpv08}
V.~N. {Zirakashvili}, V.~S. {Ptuskin}, and H.~J. {V{\"o}lk}, {\it {Modeling
  Bell's Nonresonant Cosmic-Ray Instability}},  {\em \apj} {\bf 678} (May,
  2008) 255--261, [\href{http://xxx.lanl.gov/abs/0801.4486}{{\tt
  arXiv:0801.4486}}].

\bibitem{reville+07}
B.~{Reville}, J.~G. {Kirk}, P.~{Duffy}, and S.~{O'Sullivan}, {\it {A cosmic ray
  current-driven instability in partially ionised media}},  {\em \aap} {\bf
  475} (Nov., 2007) 435--439, [\href{http://xxx.lanl.gov/abs/0707.3743}{{\tt
  arXiv:0707.3743}}].

\bibitem{ohira+09}
Y.~{Ohira}, B.~{Reville}, J.~G. {Kirk}, and F.~{Takahara}, {\it
  {Two-Dimensional Particle-In-Cell Simulations of the Nonresonant,
  Cosmic-Ray-Driven Instability in Supernova Remnant Shocks}},  {\em \apj} {\bf
  698} (June, 2009) 445--450, [\href{http://xxx.lanl.gov/abs/0812.0901}{{\tt
  arXiv:0812.0901}}].

\bibitem{mario-anatoly}
M.~A. {Riquelme} and A.~{Spitkovsky}, {\it {Magnetic Amplification by
  Magnetized Cosmic Rays in Supernova Remnant Shocks}},  {\em \apj} {\bf 717}
  (July, 2010) 1054--1066, [\href{http://xxx.lanl.gov/abs/0912.4990}{{\tt
  arXiv:0912.4990}}].

\bibitem{bac07}
P.~{Blasi}, E.~{Amato}, and D.~{Caprioli}, {\it {The maximum momentum of
  particles accelerated at cosmic ray modified shocks}},  {\em MNRAS} {\bf 375}
  (Mar., 2007) 1471--1478,
  [\href{http://xxx.lanl.gov/abs/astro-ph/0612424}{{\tt astro-ph/0612424}}].

\bibitem{long+03}
K.~S. {Long}, S.~P. {Reynolds}, J.~C. {Raymond}, P.~F. {Winkler}, K.~K. {Dyer},
  and R.~{Petre}, {\it {Chandra CCD Imagery of the Northeast and Northwest
  Limbs of SN 1006}},  {\em \apj} {\bf 586} (Apr., 2003) 1162--1178.

\bibitem{morlino+10}
G.~{Morlino}, E.~{Amato}, P.~{Blasi}, and D.~{Caprioli}, {\it {Spatial
  structure of X-ray filaments in SN 1006}},  {\em \mnras} {\bf 405} (June,
  2010) L21--L25, [\href{http://xxx.lanl.gov/abs/0912.2972}{{\tt
  arXiv:0912.2972}}].

\bibitem{jumpkin}
D.~{Caprioli}, P.~{Blasi}, E.~{Amato}, and M.~{Vietri}, {\it {Dynamical
  feedback of self-generated magnetic fields in cosmic ray modified shocks}},
  {\em \mnras} {\bf 395} (May, 2009) 895--906,
  [\href{http://xxx.lanl.gov/abs/0807.4261}{{\tt arXiv:0807.4261}}].

\bibitem{hillas05}
A.~M. {Hillas}, {\it {TOPICAL REVIEW: Can diffusive shock acceleration in
  supernova remnants account for high-energy galactic cosmic rays?}},  {\em
  Journal of Physics G Nuclear Physics} {\bf 31} (May, 2005) 95--+.

\bibitem{warren+05}
J.~S. {Warren et al.}, {\it {Cosmic-Ray Acceleration at the Forward Shock in
  Tycho's Supernova Remnant: Evidence from Chandra X-Ray Observations}},  {\em
  Ap. J.} {\bf 634} (Nov., 2005) 376--389,
  [\href{http://xxx.lanl.gov/abs/astro-ph/0507478}{{\tt astro-ph/0507478}}].

\bibitem{jumpl}
D.~{Caprioli}, P.~{Blasi}, E.~{Amato}, and M.~{Vietri}, {\it {Dynamical Effects
  of Self-Generated Magnetic Fields in Cosmic-Ray-modified Shocks}},  {\em
  \apjl} {\bf 679} (June, 2008a) L139--L142,
  [\href{http://xxx.lanl.gov/abs/0804.2884}{{\tt arXiv:0804.2884}}].

\bibitem{veb07}
A.~{Vladimirov}, D.~C. {Ellison}, and A.~{Bykov}, {\it {Nonlinear Diffusive
  Shock Acceleration with Magnetic Field Amplification}},  {\em Ap. J.} {\bf
  652} (Dec., 2006) 1246--1258,
  [\href{http://xxx.lanl.gov/abs/astro-ph/0606433}{{\tt astro-ph/0606433}}].

\bibitem{zp08b}
V.~N. {Zirakashvili} and V.~S. {Ptuskin}, {\it {The influence of the
  Alfv$\backslash$'enic drift on the shape of cosmic ray spectra in SNRs}},
  {\em astro-ph/0807.2754} (July, 2008b)
  [\href{http://xxx.lanl.gov/abs/0807.2754}{{\tt arXiv:0807.2754}}].

\bibitem{kang-ryu10}
H.~{Kang} and D.~{Ryu}, {\it {Diffusive Shock Acceleration in Test-particle
  Regime}},  {\em \apj} {\bf 721} (Sept., 2010) 886--892,
  [\href{http://xxx.lanl.gov/abs/1008.0429}{{\tt arXiv:1008.0429}}].

\bibitem{chevaliang89}
R.~A. {Chevalier} and E.~P. {Liang}, {\it {The interaction of supernovae with
  circumstellar bubbles}},  {\em \apj} {\bf 344} (Sept., 1989) 332--340.

\bibitem{garcia-low1}
G.~{Garcia-Segura} and M.~{Mac Low}, {\it {Wolf-Rayet Bubbles. I. Analytic
  Solutions}},  {\em \apj} {\bf 455} (Dec., 1995) 145--+.

\bibitem{garcia-low2}
G.~{Garcia-Segura} and M.~{Mac Low}, {\it {Wolf-Rayet Bubbles. II. Gasdynamical
  Simulations}},  {\em \apj} {\bf 455} (Dec., 1995) 160--+.

\bibitem{weaver+77}
R.~{Weaver}, R.~{McCray}, J.~{Castor}, P.~{Shapiro}, and R.~{Moore}, {\it
  {Interstellar bubbles. II - Structure and evolution \emph{[Erratum: }{\bf
  220} \apj~\emph{(1978) 742]}}},  {\em \apj} {\bf 218} (Dec., 1977) 377--395.

\bibitem{TMK99}
J.~K. {Truelove} and C.~F. {Mc Kee}, {\it {Evolution of non-radiative SNRs
  \emph{[Erratum: }{\bf 128} ApJS~\emph{(2000) 403]}}},  {\em \apj~Supplement
  Series} {\bf 120} (Feb., 1999) 299--326.

\bibitem{pz05}
V.~S. {Ptuskin} and V.~N. {Zirakashvili}, {\it {On the spectrum of high-energy
  cosmic rays produced by supernova remnants in the presence of strong
  cosmic-ray streaming instability and wave dissipation}},  {\em \aap} {\bf
  429} (Jan., 2005) 755--765,
  [\href{http://xxx.lanl.gov/abs/astro-ph/0408025}{{\tt astro-ph/0408025}}].

\bibitem{omy10}
Y.~{Ohira}, K.~{Murase}, and R.~{Yamazaki}, {\it {Escape-limited model of
  cosmic-ray acceleration revisited}},  {\em \aap} {\bf 513} (Apr., 2010) A17+,
  [\href{http://xxx.lanl.gov/abs/0910.3449}{{\tt arXiv:0910.3449}}].

\bibitem{escape}
D.~{Caprioli}, P.~{Blasi}, and E.~{Amato}, {\it {On the escape of particles
  from cosmic ray modified shocks}},  {\em \mnras} {\bf 396} (July, 2009)
  2065--2073, [\href{http://xxx.lanl.gov/abs/0807.4259}{{\tt
  arXiv:0807.4259}}].

\bibitem{EDB04}
D.~C. {Ellison}, A.~{Decourchelle}, and J.~{Ballet}, {\it {Hydrodynamic
  simulation of supernova remnants including efficient particle acceleration}},
   {\em A$\&$A} {\bf 413} (Jan., 2004) 189--201,
  [\href{http://xxx.lanl.gov/abs/astro-ph/0308308}{{\tt astro-ph/0308308}}].

\bibitem{bandiera-petruk04}
R.~{Bandiera} and O.~{Petruk}, {\it {Analytic solutions for the evolution of
  radiative supernova remnants}},  {\em \aap} {\bf 419} (May, 2004) 419--423,
  [\href{http://xxx.lanl.gov/abs/astro-ph/0402598}{{\tt astro-ph/0402598}}].

\bibitem{hl05}
J.~C. {Higdon} and R.~E. {Lingenfelter}, {\it {OB Associations,
  Supernova-generated Superbubbles, and the Source of Cosmic Rays}},  {\em
  \apj} {\bf 628} (Aug., 2005) 738--749.

\bibitem{Texas10}
D.~{Caprioli}, {\it {Supernova remnants as cosmic ray factories}},  {\em ArXiv
  e-prints} (Mar., 2011) [\href{http://xxx.lanl.gov/abs/1103.4798}{{\tt
  arXiv:1103.4798}}].

\bibitem{lagage-cesarsky83b}
P.~O. {Lagage} and C.~J. {Cesarsky}, {\it {The maximum energy of cosmic rays
  accelerated by supernova shocks}},  {\em A\&A} {\bf 125} (Sept., {1983b})
  249--257.

\bibitem{lagage-cesarsky83a}
P.~O. {Lagage} and C.~J. {Cesarsky}, {\it {Cosmic-ray shock acceleration in the
  presence of self-excited waves}},  {\em A\&A} {\bf 118} (Feb., {1983a})
  223--228.

\bibitem{pz03}
V.~S. {Ptuskin} and V.~N. {Zirakashvili}, {\it {Limits on diffusive shock
  acceleration in supernova remnants in the presence of cosmic-ray streaming
  instability and wave dissipation}},  {\em \aap} {\bf 403} (May, 2003) 1--10,
  [\href{http://xxx.lanl.gov/abs/astro-ph/0302053}{{\tt astro-ph/0302053}}].

\bibitem{mde97}
J.~{Meyer}, L.~O. {Drury}, and D.~C. {Ellison}, {\it {Galactic Cosmic Rays from
  Supernova Remnants. I. A Cosmic-Ray Composition Controlled by Volatility and
  Mass-to-Charge Ratio}},  {\em \apj} {\bf 487} (Sept., 1997) 182--+,
  [\href{http://xxx.lanl.gov/abs/astro-ph/9704267}{{\tt astro-ph/9704267}}].

\bibitem{butt09}
Y.~{Butt}, {\it {Beyond the myth of the supernova-remnant origin of cosmic
  rays}},  {\em Nature} {\bf 460} (Aug., 2009) 701--704,
  [\href{http://xxx.lanl.gov/abs/1009.3664}{{\tt arXiv:1009.3664}}].

\bibitem{eb07}
D.~C. {Ellison}, D.~J. {Patnaude}, P.~{Slane}, P.~{Blasi}, and S.~{Gabici},
  {\it {Particle Acceleration in Supernova Remnants and the Production of
  Thermal and Nonthermal Radiation}},  {\em \apj} {\bf 661} (June, 2007)
  879--891, [\href{http://xxx.lanl.gov/abs/astro-ph/0702674}{{\tt
  astro-ph/0702674}}].

\bibitem{omy11}
Y.~{Ohira}, K.~{Murase}, and R.~{Yamazaki}, {\it {Gamma-rays from molecular
  clouds illuminated by cosmic rays escaping from interacting supernova
  remnants}},  {\em \mnras} {\bf 410} (Jan., 2011) 1577--1582,
  [\href{http://xxx.lanl.gov/abs/1007.4869}{{\tt arXiv:1007.4869}}].

\bibitem{veb06}
A.~{Vladimirov}, D.~C. {Ellison}, and A.~{Bykov}, {\it {Nonlinear Diffusive
  Shock Acceleration with Magnetic Field Amplification}},  {\em \apj} {\bf 652}
  (Dec., 2006) 1246--1258,
  [\href{http://xxx.lanl.gov/abs/astro-ph/0606433}{{\tt astro-ph/0606433}}].

\bibitem{gab07}
S.~{Gabici}, F.~A. {Aharonian}, and P.~{Blasi}, {\it {Gamma rays from molecular
  clouds}},  {\em \apss} {\bf 309} (June, 2007) 365--371,
  [\href{http://xxx.lanl.gov/abs/astro-ph/0610032}{{\tt astro-ph/0610032}}].

\bibitem{lke08}
S.-H. {Lee}, T.~{Kamae}, and D.~C. {Ellison}, {\it {Three-dimensional Model of
  Broadband Emission from Supernova Remnants Undergoing Nonlinear Diffusive
  Shock Acceleration}},  {\em \apj} {\bf 686} (Oct., 2008) 325--336,
  [\href{http://xxx.lanl.gov/abs/0806.4041}{{\tt arXiv:0806.4041}}].

\bibitem{gac09}
S.~{Gabici}, F.~A. {Aharonian}, and S.~{Casanova}, {\it {Broad-band non-thermal
  emission from molecular clouds illuminated by cosmic rays from nearby
  supernova remnants}},  {\em \mnras} {\bf 396} (July, 2009) 1629--1639,
  [\href{http://xxx.lanl.gov/abs/0901.4549}{{\tt arXiv:0901.4549}}].

\bibitem{drury10}
L.~{O'C.~Drury}, {\it {Escaping the accelerator; how, when and in what numbers
  do cosmic rays get out of supernova remnants?}},  {\em ArXiv e-prints}
  (Sept., 2010) [\href{http://xxx.lanl.gov/abs/1009.4799}{{\tt
  arXiv:1009.4799}}].

\bibitem{crspectrum}
D.~{Caprioli}, E.~{Amato}, and P.~{Blasi}, {\it {The contribution of supernova
  remnants to the galactic cosmic ray spectrum}},  {\em APh} {\bf 33} (Apr.,
  2010) 160--168, [\href{http://xxx.lanl.gov/abs/0912.2964}{{\tt
  arXiv:0912.2964}}].

\bibitem{mori09}
M.~{Mori}, {\it {Nuclear enhancement factor in calculation of Galactic diffuse
  gamma-rays: A new estimate with DPMJET-3}},  {\em APh} {\bf 31} (June, 2009)
  341--343, [\href{http://xxx.lanl.gov/abs/0903.3260}{{\tt arXiv:0903.3260}}].

\bibitem{nuclei}
D.~{Caprioli}, P.~{Blasi}, and E.~{Amato}, {\it {Non-linear diffusive
  acceleration of heavy nuclei in supernova remnant shocks}},  {\em APh} {\bf
  34} (Jan., 2011) 447--456, [\href{http://xxx.lanl.gov/abs/1007.1925}{{\tt
  arXiv:1007.1925}}].

\end{thebibliography}\endgroup
\end{document}